%% file: jsacRDL_final.tex
\documentclass[journal]{IEEEtran}

\def\Figs{figs} 

\input{imports-and-defs_final}
\begin{document}
\title{Optimal Rate-Distortion-Leakage Tradeoff\\for Single-Server Information Retrieval} 

\IEEEoverridecommandlockouts

\author{Yauhen~Yakimenka, 
	    Hsuan-Yin~Lin,~\IEEEmembership{Senior Member,~IEEE,}
	    Eirik~Rosnes,~\IEEEmembership{Senior Member,~IEEE,}
	    and~J{\"o}rg~Kliewer,~\IEEEmembership{Senior Member,~IEEE}
\thanks{This work was supported partially by US National Science Foundation awards CNS-1815322 and CNS-1526547. This paper was  presented in part at the IEEE International Symposium on Information Theory (ISIT), Melbourne, Victoria, Australia, July 2021~\cite{YakimenkaLinRosnesKliewer21_1}.}
\thanks{Y. Yakimenka, H.-Y. Lin, and E. Rosnes are with Simula UiB, N-5006 Bergen, Norway (e-mail: yauhen@simula.no, lin@simula.no, eirikrosnes@simula.no).}
\thanks{J. Kliewer is with Helen and John C. Hartmann Department of Electrical and Computer Engineering, New Jersey Institute of Technology, Newark, New Jersey 07102, USA (e-mail: jkliewer@njit.edu).}}

\maketitle

\begin{abstract}
  Private information retrieval protocols guarantee that a user can \emph{privately} and \emph{losslessly} retrieve a single file from a database stored across multiple servers. In this work, we propose to simultaneously relax the conditions of perfect retrievability and  privacy in order to obtain improved download rates when all files are stored uncoded on a single server. Information leakage is measured in terms of the average success probability for the server of correctly guessing the identity of the desired file. The main findings are: i) The derivation of the optimal tradeoff between download rate, distortion, and information leakage when the file size is \emph{infinite}. Closed-form expressions of the optimal tradeoff for the special cases of ``no-leakage'' and ``no-privacy'' are also given. ii) A novel approach based on linear programming  (LP) to construct schemes for a finite file size and an arbitrary number of files. The proposed LP approach can be leveraged to find provably optimal schemes with corresponding closed-form expressions for the rate-distortion-leakage tradeoff  when the database contains at most four bits. 
%
%
Finally, for a database that contains $320$ bits, we compare  two construction methods based on the LP approach with a nonconstructive scheme downloading   subsets of files using a finite-length lossy compressor based on random coding.
\end{abstract}

\begin{IEEEkeywords}
	capacity, information leakage, information-theoretic privacy, lossy compression, private information retrieval, rate-distortion theory, single server.
\end{IEEEkeywords}


\section{Introduction}
\label{sec:introduction}


\IEEEPARstart{P}{rivate} information retrieval (PIR) was introduced in a seminal work by Chor \emph{et al.}~\cite{ChorGoldreichKushilevitzSudan95_1,ChorGoldreichKushilevitzSudan98_1} about 25 years ago. Since then, 
PIR has received a great deal of attention in both the information theory literature~\cite{ChanHoYamamoto15_1,SunJafar17_1,Freij-HollantiGnilkeHollantiKarpuk17_1, BanawanUlukus18_1,TajeddineGnilkeElRouayheb18_1,KumarLinRosnesGraellAmat19_1} (and references therein) and in the computer science literature~\cite{BeimelIshaiKushilevitRaymond02_1,Yekhanin10_1,DvirGopi16_1,CorriganGibbsKogan20_1}.
The main motivation for introducing the concept of PIR is to allow a user to download an arbitrary data item (either a single bit or a file) stored in a set of databases without disclosing in an information-theoretic sense any information about which file the user is interested in to the servers storing the databases. From the original work of Chor \emph{et al.}, it is known that downloading the entire database is optimal in terms of the overall number of bits uploaded and downloaded (or the overall communication cost) when the database is stored on a single server, hereafter referred to as the single server scenario. In order to circumvent this drawback, the early works on PIR either focused on increasing the storage overhead by replicating the database on several servers, or by relaxing the information-theoretic full privacy condition by allowing for \emph{computational} privacy, leading to the concept of  computationally-private information retrieval (CPIR). In CPIR, proposed as early as 1997 \cite{KushilevitzOstrovsky97_1},  the privacy  relies on an intractability assumption (e.g., the hardness of deciding quadratic residuosity is considered in \cite{KushilevitzOstrovsky97_1}), and in contrast to information-theoretic PIR, the identity of the requested file  can be determined precisely given enough computational resources. In 2017 \cite{LipmaaPavlyk17_1}, a rather simple CPIR protocol was proposed that brings  the communication complexity  
   down to a constant independent of the number of data items, and where the security relies on a fully homomorphic encryption scheme. 

The concept of PIR caught the attention of the information theory society about a decade ago and due to practical considerations\textemdash the noncolluding assumption being questionable in many real-world scenarios\textemdash the single server scenario has received renewed interest, e.g., by considering that the user has some priori side information on the content of the database. Furthermore, the upload cost is commonly neglected as for practical use cases the data items are reasonably big files,  and then the download cost dominates. This has led to the definition of PIR capacity, which is defined as the highest achievable download rate over all possible PIR protocols.  As shown in \cite{KadheGarciaHeidarzadehElRouayhebSprintson20_1}, the download rate can indeed be improved by considering side information, while preserving information-theoretic privacy. Two cases are considered in \cite{KadheGarciaHeidarzadehElRouayhebSprintson20_1}, namely whether or not the privacy of the side information needs to be preserved. Several parallel and follow-up works have appeared recently, see, e.g., \cite{LiGastpar18_1,HeidarzadehKazemiSprintson21_1}, and references therein. 
An alternative line of research has considered relaxing the full privacy condition of PIR in order to improve the download cost, referred to as weakly-private information retrieval (WPIR) \cite{LinKumarRosnesGraellAmatYaakobi19_1,LinKumarRosnesGraellAmatYaakobi21_1,SamyTandonLazos19_1,SamyAttiaTandonLazos21_1}. 
In~\cite{LinKumarRosnesGraellAmatYaakobi21_1}, an exact expression for the WPIR capacity in the single server scenario was derived using both mutual information and maximal leakage (MaxL) \cite{Smith09_1,IssaWagnerKamath20_1} as a privacy metric. Recently, the multi-server WPIR problem under the MaxL metric has also been studied in \cite{LinKumarRosnesGraellAmatYaakobi21_2app,ZhouGuoTian20_1}.




In this paper, in addition to relaxing the full privacy condition of PIR (as in \cite{LinKumarRosnesGraellAmatYaakobi21_1}), 
we further propose to 
relax the condition of perfect retrievability, which we  refer to as lossy WPIR (LWPIR),  
in order to obtain improved download rates compared to single-server WPIR. To the best of our knowledge, relaxing the perfect reconstruction condition in information-theoretic privacy-related problems has not been considered in the literature previously.
The motivation for allowing for some level of distortion when reconstructing the desired file at the user is that in several scenarios, e.g., when retrieving video, audio, or image files a small level of distortion is perfectly acceptable. In practice,  the range of acceptable distortion is typically limited and will be decided by the application and the user. Moreover, as is common for information-theoretic PIR, the upload cost will be ignored as typically it does not scale with the file size since queries for a small file size can be reused for larger file sizes \cite{ChanHoYamamoto15_1,SunJafar17_1}. In this work, the information leakage is measured in terms of the average success probability for the server of correctly guessing the identity of the requested file, which can be shown to be equivalent to the MaxL metric.
%

Our main contributions are two-fold:
\begin{enumerate}[nosep,label=\roman*)]
	\item The characterization of the \emph{optimal} tradeoff between download rate, distortion, and information leakage when the file size is \emph{infinite}, revealing a connection to conditional rate-distortion theory~\cite{Gray72_1,Gray73_1} (\cref{thm:minimum-download_SSLWPIR,cor:SSLWPIR-download-expression_PMFs}).
	\label{it:finding-tradeoff}
	
	\item A novel approach based on linear programming (LP) to construct schemes  for a finite file size and an arbitrary number of files.
	\label{it:finding-LP}
\end{enumerate}

 For \ref{it:finding-tradeoff}, we also show that for the special cases of ``no-leakage'' and ``no-privacy''  the optimal tradeoff can  simply be expressed as the number of files  times the conventional rate-distortion function and as the rate-distortion function by itself, respectively (\cref{cor:2-special-cases}). However, for intermediate privacy levels, as opposed to the special cases above, applying an optimal lossy compressor on top of the scheme from \cite{LinKumarRosnesGraellAmatYaakobi21_1} (as formulated in Theorem~\ref{thm:download-rate_WPIR_plus_compression}) is \emph{not} optimal (in terms of download rate) as we show in \cref{{sec:example-continued_binaryM3Q4},sec:example_binaryM3Q2_infnite-distotion}, even for the infinite file size case. Hence, there is more to the concept of LWPIR than just a straightforward  composition of WPIR and optimal lossy compression.

For \ref{it:finding-LP}, the LP approach 
is used to find provably optimal schemes  with corresponding closed-form expressions for the  rate-distortion-leakage tradeoff when the database contains two or four bits (\cref{thm:opt-tradeoff-M2-beta2,thm:opt-tradeoff-M2-beta1}). These optimal schemes can again
be used to construct schemes for a larger number of files and for a
larger file size. In a similar manner and by leveraging the LP approach, we construct schemes from lossy compressors found by a simulated annealing based heuristic search.  
For a database containing $320$ bits, we compare the two construction methods above based on the LP approach with a nonconstructive scheme based on downloading subsets of files using a finite-length lossy compressor based on random coding that is adapted from \cite[Cor.~17]{KostinaVerdu12_1}.

A similar setup to the one in this paper was considered in our companion paper \cite{WengYakimenkaLinRosnesKliewer20_1sub}. However, in  \cite{WengYakimenkaLinRosnesKliewer20_1sub} the focus was on real-world datasets and on learning efficient schemes through a generative adversarial approach. This is in  contrast to the current work which considers the case of independent and identically distributed (i.i.d.) files and the derivation of results on the \emph{optimal} rate-distortion-leakage tradeoff.



The remainder of this paper is organized as follows. Section~\ref{sec:preliminaires} presents the notation, definitions, and problem statement for single-server LWPIR. In \cref{sec:WPIR-with-LC}, we consider applying an optimal lossy compressor on top of a WPIR scheme, both for a finite and an infinite file size. In \cref{sec:optimal-tradeoff-infinite-beta}, we derive the optimal tradeoff between download rate, distortion, and information leakage for an infinite file size. The optimal tradeoff is formulated as an optimization problem that can be numerically solved. In \cref{sec:optimal-tradeoff-finite-beta}, we consider the finite file size case  and present an equivalent LP formulation. Closed-form expressions for the optimal rate-distortion-leakage tradeoff when the database contains two or four bits are also provided, along with explicit schemes for achieving these.
In \cref{sec:finite-size-LWPIR}, two construction methods of LWPIR schemes for a finite file size and binary data are suggested; 
   both based on the proposed LP approach (\cref{sec:LWPIR_from_smaller_ones,sec:rates-from-lossy-compressors}). By borrowing some useful results from finite-length lossy compression (\cref{sec:KV-lossy-compressors}) and by constructing a pool of lossy compressors by running a simulated annealing based search  (\cref{sec:simulate-annealing-search}), the two construction methods, along with the composition scheme of \cref{sec:WPIR-with-LC}, are compared for a scenario of $16$ files, each composed of $20$ bits. Finally, Section~\ref{sec:conclusion} concludes the paper. 

\section{Preliminaries and Problem Statement}
\label{sec:preliminaires}

\subsection{Notation}
\label{sec:notation}

We denote random variables (RVs) and functions by capital letters (the distinction should be clear from the context), e.g., $X$, and vectors by bold italic font, e.g., $\vec x$. 
Bold capitals denote random vectors, e.g., $\vec X$,  
calligraphic capitals denote sets, e.g., $\set X$, and sans-serif capitals denote matrices, e.g., $\mat{A}$. For a given index set $\set{N}$, we use $X^{\set{N}}$ to represent $\bigl\{X^{(m)}\colon m\in\set{N}\bigr\}$. The all-zero vector (of some arbitrary length) is denoted by $\bm 0$.
We let $[n] \eqdef \{1,2,\dotsc, n\}$ and let $\dist_{\mathrm{H}}(x,y)$ denote the Hamming distortion measure, which equals $0$ if $x=y$, and $1$ otherwise.
The set of real numbers is denoted by $\Reals$ and the set of nonnegative real numbers is denoted by $\Reals_{\ge0}$. Furthermore, some constants are depicted by Greek letters or a special font,
e.g., $\const{X}$. The probability of the event ``$X = x$'' is denoted by $\Pr{X=x}$ and 
$\EE[X]{\cdot}$ denotes expectation with respect to  $X$ (the subscript is sometimes dropped if it is clear from the context).
 The binary entropy function
is denoted by $\Hb(\cdot)$. $P_{X}$ denotes the probability distribution of the RV $X$, and if it is clear from the context, we sometimes drop the subscript, i.e., $P_{X}(\cdot)=P(\cdot)$. We write $X \sim \unif{\set X}$ to denote that $X$ is uniformly distributed over the set $\set X$ and also write $X \sim P(\cdot)$ to denote that $X$ is distributed according to $P(\cdot)$.  If $Y$ is a deterministic function of RVs $X_1, \dotsc, X_n$, we conventionally use $Y$ to denote both the function and the RV, i.e.,  $Y = Y(X_1, \dotsc, X_n)$. $\eMI{X}{Y}$ denotes the mutual information between $X$ and $Y$ (in bits).

Values of a discrete RV $X$ can be encoded into variable-length binary
codewords by an \emph{optimal} lossless source code (e.g., a Huffman
code). Throughout this paper, for every message $x$, we denote by
$\ell(x)$ the codeword length in bits,
and by $\ell^\ast(x)$ the codeword length in an optimal code. 
Furthermore, $\conv{f}{g}(x)$ represents the lower convex envelope of two functions $f(x)$ and $g(x)$ defined on the same interval $\set{I}$, i.e., $\conv{f}{h}(x)\eqdef\sup\{h\colon h \textnormal{ is convex and } h \leq \min\{f,g\} \textnormal{ over } \set{I}\}$.

\subsection{System Model}
\label{sec:system-model}

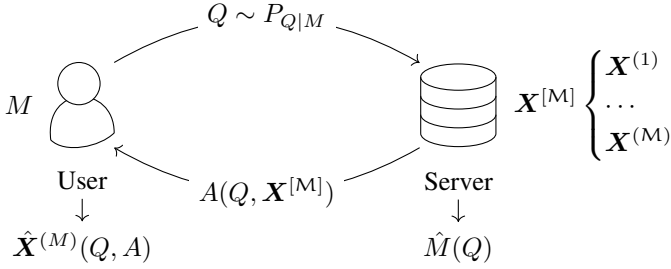
\begin{figure}
	\centering
	\input{\Figs/model.tikz}
	\vspace{-3mm}
	\caption{System model.}
	\label{fig:model}
\end{figure}

We consider the case of a single server storing $\const M$ files $\vec X^{(1)}, \dotsc, \vec X^{(\const M)}$, each of $\beta$ symbols from $\set X$, where $\vec X^{(m)} = (X_1^{(m)}, \dotsc, X_{\beta}^{(m)})$, for $m \in [\const M]$. 
We assume the files are independent and uniformly distributed over $\set X^{\beta}$. As a shorthand, we denote all the files together as $\vect{X}^{[\const{M}]} = (\vec X^{(1)}, \dotsc, \vec X^{(\const M)}) \sim \bigUniform{\set X^{\const{M}\beta}}$. The user wants to obtain the file with index $M \sim \eUniform{[\const M]}$ while keeping $M$ to a certain extent private.\footnote{Throughout the paper, we assume for simplicity that $M \sim \eUniform{[\const M]}$ and also that the file sizes are  equal and fixed. The distribution $P_{M}$ does not affect the generality of our results, and it is also common to have equal and fixed file sizes in the PIR/WPIR literature. These assumptions can be lifted, which is referred to as semantic PIR in the literature \cite{VithanaBanawanUlukus21_1app}.} Accordingly, the user generates a randomized query $Q \in \set Q$, for some set $\set Q = \{q_1,\ldots,q_{|\mathcal{Q}|}\}$, according to a conditional distribution $P_{Q|M}(q|m)$ and sends it to the server. The conditional probabilities $P(q | m)$ are considered to be public.\footnote{Similarly, we can say that the \emph{joint} probabilities $P_{M,Q} = P_{M}P_{Q|M}=\nicefrac{P_{Q|M}}{\const M}$ are public. This allows for a straightforward generalization for the case when the index $M$ is not uniformly distributed.}
Based on the query $Q$ and the files $\vec X^{[\const{M}]}$, the server produces the response $A = A(Q, \vec X^{[\const{M}]}) \in \set A$, for some set $\set A$, and sends it back to the user. Finally, the user produces an estimate of the desired file $\hat{\vec X}^{(M)} = \hat{\vec X}^{(M)}(Q, A)$, where $\hat{\vec X}^{(M)} = (\hat X_1^{(M)}, \dotsc, \hat X_{\beta}^{(M)}) \in \hat{\set X}^{\beta}$, for some set $\hat{\set X}$. 
The server produces its own guess $\hat M = \hat M(Q)  \in [\const M]$ of the index $M$. The overall system model is depicted in \cref{fig:model}.

We call $\{ \set X, \hat{\set X}, \set Q, \set A, \const M, \beta, \{P(q|m)\}, A(\cdot), \{\hat{\vec X}^{(m)}(\cdot)\} \}$ an LWPIR scheme. 
The function $\hat M(\cdot)$ is not part of the scheme as it can be chosen by the server freely. In the rest of the paper, we assume $\set X = \hat{\set X}$ is finite. Since for each fixed $Q=q$ (known both to the user and the server), $A(\cdot)$ is a deterministic function from $\set X^{\const M \beta}$ to $\set A$, without loss of generality, we can assume that $\set A$ has at most $|\set X|^{\const M \beta}$ elements, e.g., $\set A = \{1, 2, \dotsc, |\set X|^{\const M \beta} \}$. The user wants to retrieve the file $\vec X^{(M)}$ while leaking only partial information about the index $M$. 
%
The server is assumed to be \emph{honest-but-curious}, i.e., it serves the user's requests correctly but tries to learn from them what the user is interested in.

A straightforward approach is to download all the files, yet the user wants to minimize the size of the downloaded data. On the other hand, some degree of imprecision with the data is often allowed, which can potentially improve other parameters. This results in a trifold tradeoff between:
1) the download rate, 2) the distortion between the stored data and the reconstructed data  by the user, and 3) the amount of information leaked about what the user wants to download.

\subsection{Download Rate, Distortion, and Information Leakage}
\label{sec:definitons_rate-distortion-leakage}

In order to transmit the response $A$, we encode it with a lossless source code, which in general depends on the query. 
We define the \emph{download rate} as
\begin{IEEEeqnarray*}{rCl}
  R & \eqdef &\frac{\eEcond[A,Q]{\ell(A)}{Q}}{\beta}
  \\
  & = &\frac{1}{\beta} \sum_{q\in\set{Q}}P_Q(q)\eEcond[A\mid Q=q]{\ell(A)}{Q=q}.
\end{IEEEeqnarray*}


The \emph{distortion} of the user's reconstruction is defined as
\begin{align*} 
D = \frac{1}{\beta} \sum_{i=1}^{\beta} \E[M,Q,\vec X^{[\const M]}]{\dist \left(X^{(M)}_i, \hat X_i^{(M)}\right)} = \bigE[M]{D^{(M)}},
\end{align*}
where $\dist\colon \set X \times \hat{\set X} \to \Reals_{\ge0}$ is a per-symbol distortion function, which is chosen based on the particular type of data considered, and 
\[
D^{(m)} \eqdef \frac{1}{\beta} \sum_{i=1}^{\beta} \EE[Q, \vec X^{[\const M]}]{\dist \left(X^{(m)}_i, \hat X_i^{(m)}\right)},\quad \forall\, m \in [\const{M}].
\]
Note that a scheme might have distortions for some files that are worse than the average distortion $D$. 
However, as we show in \cref{prop:uniform-Ds} in Appendix~\ref{sec:properties_LWPIR} any LWPIR scheme can be transformed into a scheme with equal distortions for all the files.


The only source of undesirable leakage is the query $Q$, and we denote and define the \emph{leakage} of a given $P_{Q|M}$ as the probability of the maximum-likelihood (ML) guess as
\begin{IEEEeqnarray*}{c}
  L(P_{Q|M}) \eqdef 
  \frac{1}{\const M} \sum_{q \in \set Q} \max_{m \in [\const M]} P_{Q|M}(q|m).
\end{IEEEeqnarray*}
It is clear that we have $\nicefrac{1}{\const{M}} \leq L \leq 1$ under this privacy metric. More precisely, $L=\nicefrac{1}{\const M}$ corresponds to the ``no-leakage'' case where the server cannot do anything better than randomly guess the index $M$. On the other hand, $L=1$ corresponds to the ``no-privacy'' case where the server always guesses $M$ correctly. We remark here that the robust MaxL metric introduced in \cite{Smith09_1,IssaWagnerKamath20_1} is given by $\log{\bigl[\const{M}\cdot L(P_{Q|M})\bigr]}$. Hence, our leakage measure is equivalent to MaxL and has a clear operational meaning.\footnote{Other leakage metrics can also be considered, like, e.g., mutual information and worst-case information leakage, but they have a less clear operational meaning \cite{IssaWagnerKamath20_1,WagnerEckhoff18_1}.}

If $\dist_{\mathrm{H}}(x,y)$ is used as per-symbol distortion, then the  distortion $D$ is the average number of incorrectly reconstructed symbols of $\vec X^{(M)}$, and the best estimate is the per-symbol ML estimate
$\hat{\vec X}^{(m)} = (\hat X_1^{(m)}, \dotsc, \hat X^{(m)}_{\beta})$, where
\[
	\hat X^{(m)}_i(q, a) \eqdef \argmax_{y \in \set X} \BigPrcond{A = a}{Q = q, X^{(m)}_i = y}.
\]


Our goal is to characterize the minimum download rate (over all LWPIR schemes) under a given distortion constraint $D \leq \const{D}$ and a given leakage level $L \leq \const{L}$, for either a \emph{finite} or an \emph{infinite} file size $\beta$. In the finite setting of $\const{M}$ files of length $\beta$, we denote such a minimum rate by $R^{\ast}(\const D, \const L; \const M,\beta)$. For the infinite setting, we define $R^\ast(\const D, \const L;\const{M}) \eqdef \lim_{\beta \to \infty} R^{\ast}(\const D, \const L; \const M,\beta)$. For notational convenience, we sometimes omit the argument $\const{M}$ if it is contextually unambiguous. Similarly to the accustomed characteristic of the PIR problem, we define the single-server LWPIR capacity as the reciprocal of the minimum download rate $R^\ast(\const D, \const L; \const M)$.

Note that from a source coding perspective, given a particular query $Q=q$, the server produces the response $A_q(\vect{X}^{[\const{M}]})\eqdef A(q, \vect{X}^{[\const{M}]})$, which can be seen as a lossy compressor of $\vect{X}^{[\const{M}]}$. The rate of an LWPIR scheme is then the average compression rate over the distributions of $Q$ and $\vect{X}^{[\const{M}]}$.

This builds a strong connection between LWPIR and lossy compression. For instance, as $\beta \to \infty$, the server can separately compress each file $\vect X^{(1)}, \dotsc, \vect X^{(\const M)}$ with rate $\RDF_X(\const D)$, where $\RDF_X(\const D)$ is the rate-distortion function of the source $X$ (prototype RV for the symbols of the files) with distortion measure $\dist(x, \hat{x})$, i.e.,
    \begin{IEEEeqnarray*}{c}
      \RDF_{X}(\const{D})\eqdef\min_{\substack{P_{\hat{X}|X}(\hat{x}|x)\colon\\ \EE[X, \hat X]{\dist(X,\hat{X})}\leq\const{D}}}\eMI{X}{\hat{X}}
      \label{eq:def_RD-ft}
    \end{IEEEeqnarray*}%
    and send all the compressed versions together to the user, who can then recover the file of interest with distortion $\const D$. Such a scheme has leakage $\const L = \nicefrac{1}{\const M}$ and download rate $R = \const M \RDF_X(\const D)$.    
In the next section, such a composition of WPIR and (optimal) lossy compression is explored further.

%
%

\section{WPIR Composed With Lossy Compression}
\label{sec:WPIR-with-LC}

In this section, we consider the composition of the optimal WPIR scheme from \cite{LinKumarRosnesGraellAmatYaakobi21_1} with a lossy compressor. Clearly, this gives an LWPIR scheme. We refer to such schemes as \emph{WPIR+LC} schemes. To make the paper self-contained, we formally describe the steps of this scheme as follows. 
\begin{itemize}
\item The user chooses uniformly at random a subset $\set I \subset [\const M]$ of cardinality $\card{\set I} = \const N$ such that the index of interest $M$ is in $\set I$, and sends $\set I$ to the server.
\item The server concatenates the files indexed by $\set I$ into a block of $\const N \beta$ symbols, compresses it with some pre-agreed lossy compressor, and sends it back to the user.
\item The user reconstructs (with distortion) the compressed files and keeps only the desired one, $\hat{\vec X}^{(M)}$. The remaining $\const N-1$ files have been requested only to ``trick'' the server and are thus discarded. 
\end{itemize}

The rate and the distortion follow from the properties of the chosen lossy compressor, and from the server's perspective, $M$ is uniformly distributed over $\set I$, and thus the leakage is $\nicefrac{1}{\const N}$. In particular, the distortion is equal to the distortion of the compressor (because of uniformity of the query $\mathcal I$), while the rate is $\const N$ times the rate of the compressor.\footnote{Assuming the rate of the compressor is in bits per symbol.} This approach is general and works for both finite and infinite $\beta$. For infinite $\beta$, the compressor can be made optimal and the corresponding optimal rate is given by the well-known rate-distortion function. For a finite $\beta$, we denote by $\RDF_X(\const{D};\beta)$ the optimal rate of a finite-length source coding scheme with distortion constraint $\const{D}$ and block length $\beta$. In the limit as $\beta$ tends to infinity, $\RDF_X(\const{D};\beta)$ approaches the rate-distortion function, and for notational convenience, $\RDF_X(\const{D};\infty) \eqdef \RDF_X(\const{D})$. We remark that by time-sharing schemes with different values of $\const N$, we can construct LWPIR schemes for arbitrary leakage levels (not only reciprocals of integers).


\begin{theorem}
  \label{thm:download-rate_WPIR_plus_compression}
  Consider $\const{M}$ files and a fixed file size $\beta$ (finite or infinite). For any $\const{N} \in [\const{M}-1]$, $0\leq\alpha\leq 1$, distortion constraint $\const{D}$, and leakage constraint $\const{L} = \nicefrac{\alpha}{\const{N}} + \nicefrac{(1-\alpha)}{(\const{N}+1)}$, the download rate
  \begin{IEEEeqnarray*}{c}
   R_{\mathrm{WPIR+LC}}(\const D, \const L;\beta)=  (\const{N}+1-\alpha) \RDF_{X}(\const{D};\beta) \IEEEeqnarraynumspace
  \end{IEEEeqnarray*}
  is achievable.\footnote{In the following, $R_{\mathrm{WPIR+LC}}(\const D, \const L) \eqdef R_{\mathrm{WPIR+LC}}(\const D, \const L;\infty)$.}
  %
  %
\end{theorem}
\begin{IEEEproof}
  For any $\const{N}\in[\const{M}-1]$, we know that both $\const{N}\RDF_{X}(\const D;\beta)$ and $(\const{N}+1)\RDF_{X}(\const D;\beta)$ are achievable by using a source coding scheme with rate $\RDF_{X}(\const D;\beta)$ and distortion $\const{D}$. Then, it can be seen that the leakage $\const L=\nicefrac{\alpha}{\const{N}} + \nicefrac{(1-\alpha)}{(\const{N}+1)}$ and the download rate $\alpha \const{N}\RDF_{X}(\const D;\beta)+(1-\alpha)(\const{N}+1)\RDF_{X}(\const D;\beta)=(\const{N}+1-\alpha)\RDF_{X}(\const D;\beta)$ are also achievable in a time-sharing manner.
\end{IEEEproof}

As we will show below in \cref{sec:optimal-tradeoff-infinite-beta} (see Examples~1 (Continued) and 2 in \cref{sec:example-continued_binaryM3Q4,sec:example_binaryM3Q2_infnite-distotion}, respectively), the simple scheme of \cref{thm:download-rate_WPIR_plus_compression} is in general not optimal (in terms of download rate), even for an infinite file size and when the leakage is a reciprocal of an integer. Hence, there is more to the concept of LWPIR than just a straightforward  composition of WPIR and optimal lossy compression.

\subsection{Example~1: Binary Data With Hamming Distortion for $\const{M}=3$ Files} 
\label{sec:example1_binaryM3Q4}

We next present an example to illustrate Theorem~\ref{thm:download-rate_WPIR_plus_compression} in the infinite file size case. 
In this example, we assume $X\sim\eUniform{\{0,1\}}$ (uniform binary source). It is known that $\RDF_{X}(\const{D})=1-\Hb(\const{D})$, $0\leq\const{D}\leq\nicefrac{1}{2}$. Now, if we require the leakage to be $\const{L}=\nicefrac{1}{2}\cdot\nicefrac{1}{2}+(1-\nicefrac{1}{2})\nicefrac{1}{3}=\nicefrac{5}{12}$, Theorem~\ref{thm:download-rate_WPIR_plus_compression} shows that the WPIR+LC scheme gives a rate of $R_{\mathrm{WPIR+LC}}=\nicefrac{1}{2}\bigl[2\RDF_X(\const{D})\bigr]+\nicefrac{1}{2}\bigl[3\RDF_X(\const{D})\bigr]$. In particular, the scheme is constructed as follows. During a fraction $\alpha=\nicefrac{1}{2}$ of the time, the user requests any two of the three files, and the server uses a pre-agreed lossy compressor with rate $\RDF_X(\const{D})$ and distortion $\const{D}$ to compress each file separately. This gives a rate of $2\RDF_X(\const{D})$ and a leakage of $\nicefrac{1}{2}$. In the remaining time, the user requests all three files, and the server compresses them separately with distortion $\const{D}$ to achieve a rate of $3\RDF_X(\const{D})$ and a leakage of $\nicefrac{1}{3}$. Note that the WPIR+LC scheme we consider here is equivalent to using the following $P_1(q|m)$:
\begin{center}
  \begin{tabular}{cccccc}
    \toprule
    $q$ & $q_1$ & $q_2$ & $q_3$ & $q_4$
    \\
    \midrule
    $P_1(q|1)$ & $\nicefrac{1}{4}$ & $\nicefrac{1}{4}$ & $0$ & $\nicefrac{1}{2}$
    \\
    $P_1(q|2)$ & $\nicefrac{1}{4}$ & $0$      & $\nicefrac{1}{4}$ & $\nicefrac{1}{2}$
    \\
    $P_1(q|3)$ & $0$ & $\nicefrac{1}{4}$ & $\nicefrac{1}{4}$ & $\nicefrac{1}{2}$
    \\
    \bottomrule
  \end{tabular}
\end{center}
Here, $q_1$ is the query requesting   files $1$ and $2$, $q_2$ is the query requesting  files $1$ and $3$, $q_3$ is the query requesting   files $2$ and $3$, and $q_4$ is the query requesting all three files. In all cases the requested files are compressed with a pre-agreed lossy compressor that compresses each file separately with distortion $\const{D}$. At the user side, the received answer is  decompressed and then the desired file is extracted. Due to the structure of the table above, the desired file is always compressed into the answer. For instance, if $M=1$, then (according to the table above) either $q_1$, $q_2$, or $q_4$, but not $q_3$, is sent to the server. In all cases, the desired file is compressed into the answer, and hence a decompressed version of it can be extracted at the user.


Note that this scheme can be seen as a simple composition of the optimal WPIR scheme from \cite{LinKumarRosnesGraellAmatYaakobi21_1} and optimal lossy compression. However, as shown below, it is  suboptimal for the considered scenario.

\section{The Optimal Rate-Distortion-Leakage Tradeoff for an Infinite File Size}
\label{sec:optimal-tradeoff-infinite-beta}

In this section, we focus on the rate-distortion-leakage tradeoff for an infinite file size $\beta$. We first characterize the optimal tradeoff, and then give two examples.


\subsection{Minimum Download Rate}
\label{sec:capacity_LWPIR}

In this subsection, we derive the minimum download rate $R^\ast(\const D, \const L; \const M)$.\footnote{The results for minimum download rate automatically give analogous results for capacity.} We express it as a solution of an optimization problem, where the objective function is a weighted sum of rate-distortion functions.

\begin{theorem}
  \label{thm:minimum-download_SSLWPIR}
  The minimum download rate $R^\ast(\const D, \const L; \const M)$ of LWPIR with $\const M$ files is the minimum value of the  optimization problem
  \begin{IEEEeqnarray}{rCl}
    \IEEEyesnumber\label{eq:minimizing-download_SSLWPIR}
    \IEEEyessubnumber*
    \min_{P_{Q|M}} \min_{\{ \const{D}_q^{(m)} \}} & \quad & \sum_{q \in \set Q} \left( P_Q(q) \sum_{m \in [\const M]} \RDF_X(\const{D}_q^{(m)}) \right)
    \label{eq:download-obj-ft}\\
    \text{s.t.} & & \frac{1}{\const M} \sum_{q \in \set Q} \max_{m \in [\const M]} P_{Q|M}(q | m) \le \const L, \label{eq:download-L-constr}\\
    & & \frac{1}{\const M} \sum_{q \in \set Q} \sum_{m \in [\const M]} P_{Q|M}(q|m) \const{D}_q^{(m)} \le \const{D},\label{eq:download-D-constr}
  \end{IEEEeqnarray} 
with $\ecard{\set{Q}} = \const{M}+3$.\footnote{In the proof, we show  that without loss of optimality, the size of $\set{Q}$ can be made no larger than $\const{M}+3$. Here, we simply made it an equality: $\ecard{\set{Q}} = \const M+3$; since if the actual query size is indeed less than $\const{M}+3$, one would obtain $P_Q(q)=0$ for some queries in the optimal solution of~\eqref{eq:minimizing-download_SSLWPIR}. These queries will not influence either the rate, the distortion, or the leakage of the scheme.}
\end{theorem}


The detailed proof can be found in Appendix~\ref{sec:proof_thm1} and here we only sketch its main steps.

First, we show how to construct an LWPIR scheme from an optimal solution of~\eqref{eq:minimizing-download_SSLWPIR}. For a fixed $Q=q$, the server can compress each file $\vect{X}^{(m)}$ with a given distortion $\const{D}^{(m)}_q$ and the corresponding rate $\sum_{m \in [\const M]} \RDF_X\bigl(\const{D}_q^{(m)}\bigr)$ is achievable as $\beta\to\infty$. Averaging over the distribution of $Q$,  the optimal value of \eqref{eq:download-obj-ft} is achieved.

The converse part is shown by combining the standard converse proof for the rate-distortion function and the approaches of joint and conditional rate-distortion theory~\cite{Gray73_1}. Finally, the equality constraint on the size of $\set{Q}$ is proved by applying Carath{\'e}odory's theorem~\cite[Thm.~15.3.5]{CoverThomas06_1}. 

We remark here that for the special case where $\const{D}=0$ and the so-called \emph{normal distortion measure} is used~\cite{Yeung08_1}, it can be shown that the optimal LWPIR scheme from \cref{thm:minimum-download_SSLWPIR}  is the WPIR scheme presented in~\cite[Sec.~V]{LinKumarRosnesGraellAmatYaakobi21_1}.


In fact, given an arbitrary conditional distribution $P_{Q|M}$, the inner minimization over the variables $\{\const{D}_q^{(m)}\}$ in \cref{thm:minimum-download_SSLWPIR} can be solved, as stated in the following corollary. 

\begin{cor}
  \label{cor:SSLWPIR-download-expression_PMFs}
  Assume that the rate-distortion function $\RDF_X(\cdot)$ is differentiable in $\const{D}$. Then, 
  \begin{IEEEeqnarray*}{c}
    R^\ast(\const{D},\const{L}; \const M) = \min_{\substack{P_{Q|M}\colon\\ L(P_{Q|M})\leq\const{L}}}\sum_{q\in\set{Q}} \left( P_Q(q)\sum_{m\in[\const{M}]} \RDF_X\bigl(\const D^{(m)\ast}_q\bigr) \right),\IEEEeqnarraynumspace 
  \end{IEEEeqnarray*}
  where the per-file and per-query distortion values $\const{D}_q^{(m)\ast}$, $m \in [\const M]$, $q \in \set Q$, satisfy
  \begin{IEEEeqnarray*}{rCl}
      \frac{P_Q(q)}{P(m,q)}\frac{\dd \RDF_X}{\dd\const{D}}\big |_{\const{D}=\const{D}^{(m)\ast}_q}& = &\lambda,
      \\
      \forall\,m\in[\const{M}],q& \in &\set{Q}, \textnormal{such that } \const{D}_q^{(m)*}>0,
      \\
      \frac{P_Q(q)}{P(m,q)}\frac{\dd \RDF_X}{\dd\const{D}}\big |_{\const{D}=\const{D}^{(m)\ast}_q}& \geq &\lambda,
      \IEEEeqnarraynumspace\IEEEyesnumber\label{eq:KKT-condtions_distortions}\\
      \forall\,m\in[\const{M}],q& \in &\set{Q}, \textnormal{such that } \const{D}_q^{(m)*}=0.
    \end{IEEEeqnarray*}
    The Lagrange multiplier $\lambda$ must be chosen such that
    \begin{IEEEeqnarray}{c}
      \sum\limits_{m\in[\const{M}]}\sum\limits_{q\in\set{Q}}P(m,q)\const{D}^{(m)\ast}_q=\const{D}.
      \IEEEeqnarraynumspace\label{eq:constraint_distortions}
  \end{IEEEeqnarray}
\end{cor}
\begin{IEEEproof}
  The rate-distortion function $\RDF_X(\cdot)$ is nonincreasing, convex, and continuous (see, e.g., \cite[Ch.~3]{ElGamalKim11_1}). Fix a feasible $P_{Q|M}$ in the optimization problem~\eqref{eq:minimizing-download_SSLWPIR}, and thus $P_Q$ is also fixed. Then, the objective function $\sum_{q \in \set Q} (P_Q(q) \sum_{m \in [\const M]} \RDF_X(\const{D}_q^{(m)}))$ is a nonnegative weighted sum of convex functions, and therefore it is convex (cf.~\cite[Sec.~3.2.1]{BoydVandenberghe04_1}). The result then follows immediately from the Karush–Kuhn–Tucker optimality conditions for convex minimization problems~\cite[Sec.~5.5.3]{BoydVandenberghe04_1}.
\end{IEEEproof}

The following corollary gives expressions for the minimum download rate in two special cases.
\begin{cor}\label{cor:2-special-cases}
	The minimum download rate $R^\ast(\const D, \const L; \const M)$ of LWPIR with $\const M$ files in the ``no-leakage'' and ``no-privacy'' special cases are
 	$R^\ast(\const D, \const L = \nicefrac{1}{\const M}; \const M) = \const M \RDF_X(\const D)$ and 
	$R^\ast(\const D, \const L = 1; \const M) = \RDF_X(\const D)$,
	respectively.
\end{cor}

\begin{IEEEproof}
See Appendix~\ref{sec:proof-cor:2-special-case}.
\end{IEEEproof}


\subsection{Example~1 (Continued)}
\label{sec:example-continued_binaryM3Q4}

We now present a scheme by using~\cref{cor:SSLWPIR-download-expression_PMFs} to obtain the optimal rate-distortion tradeoff for Example~1 (i.e., for the same given $P_1(q|m)$, but the corresponding per-file and per-query distortions will be designed according to \cref{cor:SSLWPIR-download-expression_PMFs}).\footnote{As before, $q_j$, $j\in[3]$, denote the queries requesting any two of the three files, and $q_4$ is the query that requests all the files.} Note that $\nicefrac{\dd \RDF_X}{\dd \const{D}} = \log_2{\left(\nicefrac{\const{D}}{(1-\const{D})}\right)}$. 


From \cref{cor:SSLWPIR-download-expression_PMFs}, we get the optimal values  $\{\const{D}^{(m)\ast}_q\}_{m \in [3], q\in\{q_1,\ldots,q_4\}}$ as follows:
\begin{center}
  \begin{tabular}{cccccc}
    \toprule
    $q$ & $q_1$ & $q_2$ & $q_3$ & $q_4$
    \\
    \midrule
    $\const{D}^{(1)*}_q$ & $\const{D}^\ast_1$ & $\const{D}^\ast_1$ & $0$ & $\const{D}^\ast_2$
    \\[1mm]
    $\const{D}^{(2)*}_q$ & $\const{D}^\ast_1$ & $0$      & $\const{D}^\ast_1$ & $\const{D}^\ast_2$
    \\[1mm]
    $\const{D}^{(3)*}_q$ & $0$ & $\const{D}^\ast_1$ & $\const{D}^\ast_1$ & $\const{D}^\ast_2$
    \\[1mm]
    \bottomrule
  \end{tabular}
\end{center}
Furthermore, from \eqref{eq:KKT-condtions_distortions} and \eqref{eq:constraint_distortions}, $\const{D}^\ast_1$ and $\const{D}^\ast_2$ should satisfy
\begin{IEEEeqnarray*}{rCl}
    \frac{2}{1} \frac{\dd \RDF_X}{\dd \const{D}}\bigl(\const{D}^\ast_1\bigr)&=&\frac{3}{1}\frac{\dd \RDF_X}{\dd \const{D}}\bigl(\const{D}^\ast_2\bigr),
    \\[1mm]
    \frac{1}{2}\const{D}^\ast_1+\frac{1}{2}\const{D}^\ast_2&=&\const{D},
\end{IEEEeqnarray*}
for $\const{D}^\ast_1$, $\const{D}^\ast_2>0$. From this, the optimal solution $\const{D}_1^\ast$ is the (unique) root of 
\begin{IEEEeqnarray*}{c}
  \left(\frac{1}{2\const{D}-\const{D}_1^\ast}-1\right)^\frac{3}{2}-\frac{1}{\const{D}_1^\ast}+1,
\end{IEEEeqnarray*}
and $\const{D}_2^\ast=2\const{D}-\const{D}^\ast_1$. We remark that the optimal solution does not satisfy $\const{D}_1^\ast=\const{D}_2^\ast=\const{D}$, provided $\const D < \nicefrac{1}{2}$. This implies that an WPIR+LC scheme cannot be optimal.


\subsection{Example~2: $\const{K}$-ary Data With Hamming Distortion for $\const{M}=3$ Files}
\label{sec:example_binaryM3Q2_infnite-distotion}

\begin{figure}[t!]
  \input{\Figs/CoverThomas06_1_problem10_5.tex}
  \caption{Download rate versus distortion of LWPIR schemes for $\const{M}=3$ files, leakage $\const{L}=\nicefrac{1}{2}$, and $\mathcal{X}=[64]$ with Hamming distortion.}
  \label{fig:RDL_uniform-source_HammingD}
\end{figure}
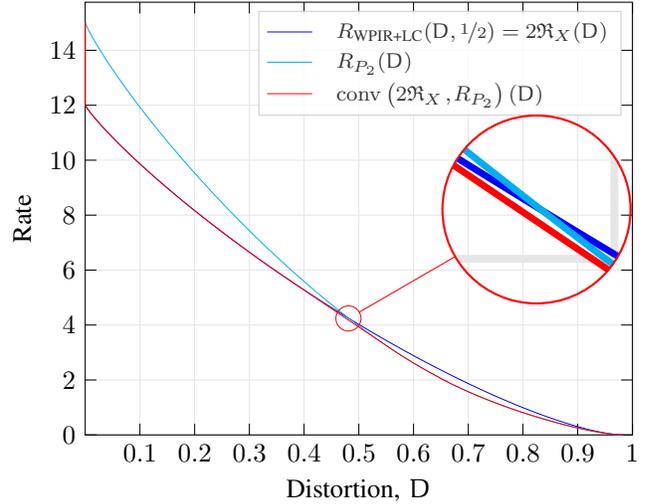

We now present another example, which shows that Theorem~\ref{thm:minimum-download_SSLWPIR} could also give a better (compared  \cref{thm:download-rate_WPIR_plus_compression}) LWPIR scheme when the leakage is a reciprocal of an integer, i.e., when $\const{L}=\frac{1}{\const{N}}$, $\const{N}\in [\const{M}]$. In this example, we consider a $\const{K}$-ary uniform source $X\sim\eUniform{[\const{K}]}$ with Hamming distortion~\cite[Problem~10.5]{CoverThomas06_1}. Its rate-distortion function can be shown to be
\begin{IEEEeqnarray*}{c}
  \RDF_X(\const{D})=\log_2{\const{K}}-\Hb(\const{D})-\const{D}\log_2{(\const{K}-1)},
\end{IEEEeqnarray*}
for $0\leq\const{D}\leq 1-\nicefrac{1}{\const{K}}$. 
Next, let $P_2(q|m)$ be defined as\footnote{Here, $q_1$ is the query requesting the first file, $q_2$ is the query requesting the second file, $q_3$ is the query requesting the third file, and $q_4$ is the query requesting all three files.} 
%
\begin{center}
  \begin{tabular}{cccccc}
    \toprule
    $q$ & $q_1$ & $q_2$ & $q_3$ & $q_4$
    \\
    \midrule
    $P_2(q|1)$ & $\nicefrac{1}{4}$ & $0$ & $0$ & $\nicefrac{3}{4}$
    \\
    $P_2(q|2)$ & $0$ & $\nicefrac{1}{4}$ & $0$ & $\nicefrac{3}{4}$
    \\
    $P_2(q|3)$ & $0$ & $0$ & $\nicefrac{1}{4}$ & $\nicefrac{3}{4}$
    \\
    \bottomrule
  \end{tabular}
\end{center}
The leakage for $P_2(q|m)$ is given by $L(P_2(q|m))=\nicefrac{1}{2}$. Applying~\cref{cor:SSLWPIR-download-expression_PMFs} we  get the optimal solution of $\{\const{D}^{(m)\ast}_q\}_{m \in [3], q\in\{q_1,q_2,q_3,q_4\}}$ as follows:
\begin{center}
  \begin{tabular}{cccccc}
    \toprule
    $q$ & $q_1$ & $q_2$ & $q_3$ & $q_4$
    \\
    \midrule
    $\const{D}^{(1)*}_q$ & $\const{D}^\ast_1$ & $0$ & $0$ & $\const{D}^\ast_2$
    \\[1mm]
    $\const{D}^{(2)*}_q$ & $0$ & $\const{D}^\ast_1$ & $0$ & $\const{D}^\ast_2$
    \\[1mm]
    $\const{D}^{(3)*}_q$ & $0$ & $0$ & $\const{D}^\ast_1$ & $\const{D}^\ast_2$
    \\[1mm]
    \bottomrule
  \end{tabular}
\end{center}
Similar to Example~1 of~\cref{sec:example-continued_binaryM3Q4}, from \eqref{eq:KKT-condtions_distortions} and \eqref{eq:constraint_distortions}, $\const{D}^\ast_1$ and $\const{D}^\ast_2$ must satisfy
\begin{IEEEeqnarray*}{rCl}
    \frac{\dd \RDF_X}{\dd \const{D}}\bigl(\const{D}^\ast_1\bigr)&=&\frac{3}{1}\frac{\dd \RDF_X}{\dd \const{D}}\bigl(\const{D}^\ast_2\bigr),
    \\[1mm]
    \frac{1}{4}\const{D}^\ast_1+\frac{3}{4}\const{D}^\ast_2&=&\const{D},
\end{IEEEeqnarray*}
for $\const{D}^\ast_1$, $\const{D}^\ast_2>0$. Since there is no apparent closed-form solution for $\const{D}_1^\ast$ and $\const{D}_2^\ast$, we instead numerically solve the system of equations for $\const{K}=64$ and plot the corresponding achievable rate curve in~\cref{fig:RDL_uniform-source_HammingD}. Here, the blue curve corresponds to $R_\textnormal{WPIR+LC}(\const{D},\nicefrac{1}{2})=2\RDF_X(\const{D})$ (the download rate from \cref{thm:download-rate_WPIR_plus_compression}), and the cyan curve represents the optimal rate-distortion curve for the given $P_2(q|m)$, where its rate is represented by $R_{P_2}(\const{D})$. As can be seen from the figure, we obtain a lower rate-distortion tradeoff curve $\conv{2\RDF_X}{R_{P_2}}(\const{D})$ than $2\RDF_X(\const{D})$ by taking the lower convex envelope of the blue and cyan curves (marked in red).





From the example above and Example~1 in \cref{sec:example-continued_binaryM3Q4},  
we observe that the optimality of an WPIR+LC scheme strongly depends on the behavior of the corresponding nonlinear rate-distortion function. In fact, given a fixed $P(q|m)$, the first condition of \eqref{eq:KKT-condtions_distortions} indicates that the tangent lines of the curves $\nicefrac{P_Q(q)\RDF_X(\const{D})}{P(m,q)}$ at the optimal point $\const{D}^{(m)\ast}_{q}$ must be parallel. Thus, a joint time-sharing approach in both leakage and distortion is necessary\textemdash a scheme directly composed of WPIR and optimal lossy compression is in general suboptimal, depending on the chosen distortion measure.

\section{The Rate-Distortion-Leakage Tradeoff for a Finite File Size}
\label{sec:optimal-tradeoff-finite-beta}

In this section, we consider the rate-distortion-leakage tradeoff for a finite file size $\beta$. We first   give an equivalent LP formulation,  and then derive the exact optimal tradeoff for $\const M \beta \leq 4$ using the LP formulation.

\subsection{Equivalent Linear Programming Formulation}

\label{sec:LP-formulation}

In this subsection, we assume that the alphabet $\mathcal{X}$ is finite and show how to transform the problem of finding $R^{\ast}(\const D, \const L; \const M,\beta)$ into an LP. The solution of the LP provably provides the value of $R^\ast(\const D, \const L; \const M,\beta)$ and a corresponding scheme.

First, we concentrate on response functions. For a fixed value of the query $Q = q$, it is a deterministic function 
\[
	A_q(\vec x): \set X^{\const M \beta} \to \set A
\]
of $\const M \beta$ symbols.  A fixed response function $A_q(\cdot)$ defines a RV $A_q(\vec X^{[\const{M}]})$ with rate
\begin{IEEEeqnarray}{c}
\label{eq:response-function-rate-def}
  R_q \eqdef \frac{\LL{\vec{X}^{[\const{M}]}}{A_q\bigl(\vec{X}^{[\const{M}]}\bigr)}}{\beta},\IEEEeqnarraynumspace
\end{IEEEeqnarray}
where, as before, $\ell^\ast(\cdot)$ stands for the codeword length in bits for an optimal lossless source code for $\vec{X}^{[\const{M}]}$.

Second, the per-file distortion is 
\begin{IEEEeqnarray*}{c}
  D^{(m)}_q \eqdef \frac 1\beta \sum_{i=1}^\beta \BigE[\vect{X}^{(m)}]{\dist \bigl(X^{(m)}_i, \hat X_i^{(m)}\bigr) \mid Q=q},\, \forall\, m \in [\const{M}].\IEEEeqnarraynumspace 
\end{IEEEeqnarray*}

An important observation is that for a fixed $Q=q$, the values $R_q$ and $D^{(m)}_q$, $m \in [\const M]$, are only determined by the function $A_q(\cdot)$ and do not depend on the way the user generates other queries. Thus, we can express the rate and distortion of an LWPIR scheme as
\begin{IEEEeqnarray*}{rCl}
	R & = & \EE[Q]{R_Q} = \frac 1{\const M} \sum_{m\in[\const{M}]} \sum_{q \in \set Q} P(q | m) R_q,\\
	D & = & \EE[M, Q]{D^{(M)}_Q} = \frac 1{\const M} \sum_{m\in[\const{M}]} \sum_{q \in \set Q} P(q | m) D^{(m)}_q,
\end{IEEEeqnarray*}
and the leakage of the ML estimate by the server as
\[
  L = \frac{1}{\const M} \sum_{q \in \set Q} \max_{m \in [\const M]} P(q | m).
\]

Now, assume all the possible response functions $\{ A_q\colon q \in \set Q \}$ defined over $\set{X}^{\const{M}\beta}$ are given, and their corresponding $R_q$, $D^{(m)}_q$, $m \in [\const M]$, are pre-calculated. Then, the problem of rate minimization can be formulated as an LP in the decision variables $\set{P}\eqdef\{ P(q|m)\colon m \in [\const M], q \in \set Q \}$ for each pair of target values of distortion $\const D$ and leakage $\const L$ as

\begin{IEEEeqnarray}{rCl}
  \IEEEyesnumber\label{eq:LP-formulation}
  \IEEEyessubnumber*
  \min_{\set{P}} & \quad & \frac{1}{\const M} \sum_{m \in [\const M]} \sum_{q \in \set Q} P(q|m) R_q
  \label{eq:LP-formulation-obj}\\
  \text{s.t.} & \quad & \sum_{q \in \set Q} P(q|m) = 1,\qquad m \in [\const M],
  \nonumber\\
  & & \frac{1}{\const M} \sum_{m \in [\const M]} \sum_{q \in \set Q} P(q|m) D_q^{(m)} \le \const{D},\label{eq:LP-formulation-D-constraint} \\
  & & \frac{1}{\const M} \sum_{q \in \set Q} \xi_q \le \const L,\label{eq:LP-formulation-L-constraint} \\
  & & 0 \le P(q|m) \le \xi_q,\qquad q \in \set Q,\, m \in [\const M].\nonumber\IEEEeqnarraynumspace
\end{IEEEeqnarray}
Here, the auxiliary variables $\xi_q$ are introduced to model the behavior of the $\max$ function.

\subsubsection{Function Filtering}\label{sec:fliter-function}
To obtain an optimal solution, one needs in principle to consider all possible $(\ecard{\set{X}}^{\const{M}\beta})^{\ecard{\set X}^{\const M \beta}}$ response functions, which grows super-exponentially in $\const M$ and $\beta$. However, many of them are in fact equivalent up to a permutation of $\set A$. More precisely, consider the following (unordered) partition of the set $\set X^{\const M \beta}$ induced by $A_q$: $\vect{x}, \vect{x}' \in \set X^{\const M \beta}$ belong to the same part of the partition if and only if $A_q(\vect{x}) = A_q(\vect{x}')$. Then if two functions $A_{q_1}$ and $A_{q_2}$ induce the same partition of $\set X^{\const M \beta}$, we call them equivalent. Intuitively, this means that there exists a bijection $\pi: \set A \to \set A$ between outputs of $A_{q_1}$ and $A_{q_2}$, i.e.,  for all $\vec x \in \set X^{\const M \beta}$, $A_{q_1}(\vec x) = \pi(A_{q_2}(\vec x))$. Equivalent functions have the same rates $R_q$ and per-file distortions $D^{(m)}_q $. Furthermore, many more functions can be filtered out as described in the following.

Associate with each $A_q$, $q \in\set{Q}$, the $(\const M+1)$-dimensional point 
$
\vec c_q = (R_q, D_q^{(1)}, D_q^{(2)}, \dotsc, D_q^{(\const M)}),
$
and define
\[
\vec c_\mathrm{max}\triangleq\Bigl(\max_{q \in \set Q} R_q, \max_{q \in \set Q} D_q^{(1)}, \max_{q \in \set Q} D_q^{(2)}, \dotsc, \max_{q \in \set Q} D_q^{(\const M)}\Bigr).
\]
Form the set of points $\set V = \{ \vec c_q\colon q \in \set Q \} \cup \{ \vec c_{\mathrm{max}}\}$ and construct its convex hull $\set P(\set V)$. Since $\set V$ is finite, $\set P(\set V)$ is a convex polytope whose vertices are points from $\set V$.
Clearly, $\vec c_{\mathrm{max}}$ is necessarily a vertex of $\set{P}(\set{V})$, as it is not less than any of the original points in each of the coordinates. Let the remaining vertices of $\set P(\set V)$ be $\bm c_{q_1}, \bm c_{q_2},\ldots$, and define the set of corresponding queries as $\set Q_{\textnormal v}\eqdef\{q_1,q_2,\ldots\} \subset \set Q$.

\begin{lemma}
  \label{lem:vertices-set_query-set}
  The optimal solution of (\ref{eq:LP-formulation}) can be obtained by choosing $\set{Q}=\set{Q}_\textnormal{v}$. 
  In other words, all $P(q|m)$ can be set to $0$ for $q \in \set Q \setminus \set{Q}_\textnormal{v}$.
\end{lemma}

\begin{IEEEproof}
See Appendix~\ref{sec:proof-lemma-1}.
\end{IEEEproof}

While the problem of finding the vertices of a convex hull can be solved (somehow) efficiently (e.g., using the Quickhull algorithm~\cite{BarberDobkinHuhdanpaa96_1}), the efficient search for candidate response functions (that would be an input for the Quickhull algorithm) in general remains an open question for future research.

Alternatively, one can restrict the response functions to a subclass, for which the values of $R_q$ and $D^{(m)}_q$ are relatively easy to calculate. Using these functions in the LP, we can find provably optimal solutions in this restricted case. The schemes found can be used as suboptimal constructive schemes for the original (unrestricted) problem.

\subsubsection{Beyond Function Filtering for Larger $\const M$ and $\beta$} \label{sec:beyond-filtering}

Assume that for one file of size $\beta$ we have $k$ response functions $A_1, A_2, \dotsc, A_k: \set X^\beta \to \set A$ having distortions $D_1, D_2, \dotsc, D_k$ and rates $R_1, R_2, \dotsc, R_k$, correspondingly.  
Now, if we want to construct schemes for $\const M$ files of size $\beta$, we can form a pool of response functions for $\const M$ files, where a response function is combined from responses for each file. More precisely, we define $\set Q = [k]^{\const M}$ and for $\vect{q} = (i_1, i_2, \dotsc, i_\const{M})\in\set Q$ and $\vect{x}^{(1)},\vect{x}^{(2)},\ldots,\vect{x}^{(\const{M})}\in\set{X}^\beta$, we have 
\[
A_{\vec{q}} (\vect{x}^{(1)}, 
\ldots,\vect{x}^{(\const{M})}) = \bigl(A_{i_1}(\vec x^{(1)}), 
\ldots, A_{i_\const{M}}(\vec x^{(\const{M})})\bigr).
\]
The rate of $A_{\vec q}$ is $R_{\vec q} = R_{i_1} + R_{i_2} + \dotsb + R_{i_\const{M}}$ and the per-file distortions are $D_{\vec q}^{(m)} = D_{i_m}$ for all $m \in [\const M]$.\footnote{In fact, the responses $A_{i_1}(\vect{x}^{(1)}), A_{i_2}(\vect{x}^{(2)}), \dotsc, A_{i_\const{M}}(\vect{x}^{(\const{M})})$ can be re-encoded together and thus  $R_{\vec q}$ is potentially a bit smaller, but we omit this for the sake of clarity.} Now, with these functions we can solve the LP in \eqref{eq:LP-formulation}.

The same idea can be used to combine response functions with multiple number of files as inputs, 
  but for the sake of clarity, we do not describe this in detail.

  \subsubsection{Solving the Optimization Problem of \cref{thm:minimum-download_SSLWPIR}}
  \label{sec:solving-opt-problem}
  
A similar LP formulation as the one in \eqref{eq:LP-formulation} can be used to solve the optimization problem in \eqref{eq:minimizing-download_SSLWPIR} of \cref{thm:minimum-download_SSLWPIR} with any required precision. For that, we find a piecewise-linear continuous approximation $\RDF^{\textrm{PWL}}_X(\const D)$ of $\RDF_X(\const D)$ defined in some $s$ approximation points $\const{D}_1 \le \dotsb \le \const{D}_s$.
If the approximation error is bounded as $\left| \RDF^{\textrm{PWL}}_X(\const D) - \RDF_X(\const D) \right| \leq \varepsilon$ for any $\const D$ and we substitute $\RDF_X(\const D)$ in \eqref{eq:minimizing-download_SSLWPIR} with $\RDF^{\textrm{PWL}}_X(\const D)$, the error in the objective function will not be larger than $\const M \varepsilon$. 

Next, assume that in the optimal solution of \eqref{eq:minimizing-download_SSLWPIR} with $\RDF_X^{\textrm{PWL}}$ we have for some $m_0$ and $q_0$ that $\const{D}_{q_0}^{(m_0)} = (1-\alpha)\const{D}_i + \alpha \const{D}_{i+1}$, $\alpha \in (0,1)$, where $\const{D}_i$ and $\const{D}_{i+1}$ are the two closest approximation points to $\const{D}_{q_0}^{(m_0)}$. Then, we can substitute the query $q_0$ with new $q'$ and $q''$, such that $\const{D}_{q'}^{(m_0)} = \const{D}_i$, $\const{D}_{q''}^{(m_0)} = \const{D}_{i+1}$, and $\const{D}_{q'}^{(m)}  = \const{D}_{q''}^{(m)} = \const{D}_{q_0}^{(m)}$ for $m \in [\const M] \setminus \{m_0\}$. Additionally, $P_{Q|M}(q'|m) = (1-\alpha)P_{Q|M}(q_0|m)$ and $P_{Q|M}(q''|m) = \alpha P_{Q|M}(q_0|m)$ for all $m \in [\const M]$. It is not difficult to check that this new scheme has exactly the same rate, distortion, and leakage.
	
Continuing in the same manner, we can construct a scheme where each of the values $\{ \const{D}_q^{(m)}\}$ belong to the set of approximation points. Thus, we can search for optimal solutions with this restriction. In other words, we can set $\set Q = [s]^{\const M}$ and for $\vec q = (i_1, \dotsc, i_{\const M}) \in \set Q$, $\const{D}_{\vec q}^{(m)} = \const{D}_{i_m}$. Thus, $\{\const{D}_{\vec q}^{(m)}\}$ are not decision variables any more and the optimization problem becomes an LP. The drawback of this approach is a possible large size of such a $\set Q$ but this is compensated by available fast LP solvers.

Finally, the problem in \eqref{eq:minimizing-download_SSLWPIR} with a piecewise-linear rate-distortion function is equivalent to the LP formulation of \eqref{eq:LP-formulation}. 

As a remark, we note that since the LWPIR problem can be reformulated as an LP, the optimal tradeoff curve $R^\ast(\const D, \const L; \const M, \beta)$ is a piecewise-linear, convex, nonincreasing function of $\const D$ when $\const L$ is fixed (or vice versa) and $\set X$ is finite. This follows since the number of response functions is finite for a finite $\set X$ (cf.~\cite[Thm.~6.6]{DantzigThapa03_1}).


\subsection{Optimal Tradeoffs for $\const{M} \beta  \leq 4$ and $\set{X} = \{0,1\}$}
\label{sec:optimal_tradeoff_M2_beta2}

In this subsection, we assume $\set{X}={\{0,1\}}$ with Hamming distortion $\dist_{\mathrm{H}}$. To illustrate the LP method in \cref{sec:LP-formulation}, we present more details for the cases of 1) $\const M=2$ files and $\beta=1$ bit, and 2) $\const M=2$ files and $\beta=2$ bits. We start with the former and state the exact optimal rate-distortion-leakage tradeoff in the following theorem.

\begin{theorem}[Optimal tradeoff for $\const M=2$, $\beta=1$
  ]
  \label{thm:opt-tradeoff-M2-beta1}
  For $\const M=2$ files each of $\beta=1$ bit, the minimum rate is
  \begin{IEEEeqnarray*}{c}
    R^\ast(\const D, \const L;\beta) =
    \begin{cases}
      3-2 \const L -4\const D &\text{ if } \const D \in \left[ 0, 1-\const L \right], \\
      1-2\const D &\text{ if } \const D \in \left[ 1-\const L, \nicefrac 12 \right].\\
    \end{cases}		
    \IEEEeqnarraynumspace
  \end{IEEEeqnarray*}
\end{theorem}

\begin{IEEEproof}
See Appendix~\ref{app:M2_beta1}.
\end{IEEEproof}

In contrast to \cref{thm:opt-tradeoff-M2-beta1}, we do not provide a formal proof  for the second case of $\const M=2$ files and $\beta=2$ bits as such a proof  does not bring much more insight compared to the proof outlined in Appendix~\ref{app:M2_beta1} and furthermore as the technical derivations become very involved. Hence, we proceed as follows.

For the latter case of $\const M=2$ files and $\beta=2$ bits, since the input to each $A_q$ is 4 bits in total, there are not more than $2^4$ different elements in its image and thus the size of $\set A$ can be limited to $2^4$. Therefore, the number of different functions from $\{0, 1\}^4$ to $\set A$ is $(2^4)^{2^4} \approx 1.8 \times 10^{19}$. Discarding all the equivalent ones, we have roughly $10^{10}$, which further drops to $11$ after filtering as outlined in \cref{sec:fliter-function}. An LP of this size can be solved easily, e.g., by using  Gurobi \cite{gurobi}. Moreover, the LP can be solved symbolically which gives a closed-form expression for the optimal tradeoff.

\begin{theorem}[Optimal tradeoff for $\const M=2$, $\beta=2$
  ] \label{thm:opt-tradeoff-M2-beta2}
  For $\const M=2$ files each of $\beta=2$ bits, the minimum rate is
  \begin{IEEEeqnarray*}{c}
    R^\ast(\const D, \const L;\beta) =
    \begin{cases}
      -\frac{11}{2} \const D + 3-2 \const L &\!\!\!\!\! \text{ if } \const D \in \left[ 0, \frac{1-L}{4} \right],
      \\[1mm]
      -5\const D + \frac{23-15\const L}{8} &\!\!\!\!\! \text{ if } \const D \in \left[ \frac{1-\const L}{4}, \frac{3(1-\const L)}{8} \right],
      \\[2mm]
      -4\const D + \frac{5-3\const L}{2} &\!\!\!\!\! \text{ if } \const D \in \left[ \frac{3(1-\const L)}8, \frac{5(1-\const L)}8 \right],
      \\[2mm]
      -\frac 83 \const D + \frac{5-2\const L}3 &\!\!\!\!\! \text{ if } \const D \in \left[ \frac{5(1-\const L)}8, 1-\const L \right],
      \\[1mm]
      -2\const D + 1 &\!\!\!\!\! \text{ if } \const D \in \left[ 1-\const L, \frac 12 \right].
    \end{cases}		
    \IEEEeqnarraynumspace
  \end{IEEEeqnarray*}
\end{theorem}

Using the same approach, we can find a closed-form expression for the optimal tradeoff for the case of  $\const M=4$ files and $\beta=1$ bit as well. However, we omit the details.

In Appendix~\ref{sec:exampl_small-optimal-schemes}, we describe explicit schemes able to meet the optimal rate-distortion-leakage tradeoff for all binary cases  with $\const M \beta = 4$.


\section{Finite-Size LWPIR Schemes}
\label{sec:finite-size-LWPIR}

In this section, we present some schemes for $\set X = \{0,1\}$ with Hamming distortion $\dist_\textnormal{H}$.

\subsection{LWPIR Schemes From Small Optimal Schemes}
\label{sec:LWPIR_from_smaller_ones}

We can further use the optimal schemes obtained for $\const M \beta \le 4$ with the LP method (see \cref{sec:optimal_tradeoff_M2_beta2}) in order to construct schemes for larger $\const M$ and $\beta$. First, the longer files can be split into smaller blocks and then a small scheme is run on the corresponding blocks. The resulting scheme has the same rate, distortion, and leakage as the small scheme. Second, a large set of files can be split into subsets and a small scheme is run on each subset of the files. The construction gives a higher rate but smaller leakage, while distortion does not change. 

We note that in both approaches the user needs to generate the query only once and can reuse it for all instances of the small scheme. Additionally, all the answers obtained from the small schemes can be re-coded together, thus potentially decreasing the overall rate. Finally, the two aforementioned constructions can be combined.

For example, consider an optimal scheme $\set S$ for $\const M=2$ files of $\beta=2$ bits each with leakage $\const L=\nicefrac{1}{2}$, distortion $\const D=\nicefrac{5}{16}$, and rate $\const R=\nicefrac{1}{2}$. The scheme uses only one response function (i.e., $\set Q = \{ q_0 \}$), which outputs $\mathtt 0$ if $\vec X^{[2]} \in$ \{1100, 1010, 0110, 1110, 1111\}, and $\mathtt 1$ otherwise.\footnote{In the codewords, we use the alphabet consisting of $\mathtt 0$ and $\mathtt 1$ in typewriter font for visual clarity.} Note that $\Pr{A = \mathtt{0}} = 1 - \Pr{A = \mathtt{1}} = \nicefrac{5}{16}$. The response $\mathtt 0$ is decoded to the estimates $(\hat{\vec X}^{(1)}, \hat{\vec X}^{(2)}) = 1110$, and $\mathtt 1$ to $0001$.

Assume we need a scheme $\set S'$ for $\const M'=2$, $\beta'=20$, and $\const L'=\nicefrac{1}{2}$. As in $\set S$, the user always sends the (only) query $q_0$. The server splits its $20$-bit files into $2$-bit blocks, and apply the same response functions to each corresponding pair of $2$-bit blocks. For example:
	\[
	\begin{array}{rcccccccccc}
		\vec X^{(1)}: & 00 & 11 & 11 & 10 & 10 & 00 & 00 & 10 & 01 & 01 \\
		\vec X^{(2)}: & 01 & 11 & 10 & 11 & 10 & 11 & 01 & 01 & 00 & 01 \\
		\hline
		A :           & \mathtt 1 & \mathtt 0 & \mathtt 0 & \mathtt 1 & \mathtt 0 & \mathtt 1 & \mathtt 1 & \mathtt 1 & \mathtt 1 & \mathtt 1 \\
	\end{array}
	\]
 The server's response is then the $10$-bit string $\mathtt{1001011111}$, which is sent back to the user. The user decodes each bit of the response into a $4$-bit block (in the order corresponding to encoding) and obtains two $20$-bit estimates $\hat{\vec X}^{(1)}$ and $\hat{\vec X}^{(1)}$:
 	\[
 \begin{array}{rcccccccccc}
 	A :           & \mathtt 1 & \mathtt 0 & \mathtt 0 & \mathtt 1 & \mathtt 0 & \mathtt 1 & \mathtt 1 & \mathtt 1 & \mathtt 1 & \mathtt 1 \\
 	\hline
 	\hat{\vec X}^{(1)}: & 00 & 11 & 11 & 00 & 11 & 00 & 00 & 00 & 00 & 00 \\
 	\hat{\vec X}^{(2)}: & 01 & 10 & 10 & 01 & 10 & 01 & 01 & 01 & 01 & 01 \\
 \end{array}
 \]
 	Finally, the user uses the estimate of the file she wanted. The rate of such a scheme is $\nicefrac{10}{20} = \nicefrac{1}{2}$. However, before sending back to the user, this $10$-bit string can be compressed using a Huffman code, which has an average length of $8.99$ bits (averaging over the distribution of $10$-bit responses). The rate of the new scheme is $\const R'=8.99/20=0.45$ and the distortion is $\const D'=\const D=\nicefrac{5}{16}$.
  
  We used the scheme with a single query for simplicity. If the size of the query set $\set Q$ would be larger than one, the user would generate one of the queries and send it to the server. The further procedure will be similar just with a different response function (and the decoding procedure corresponding to that response function).

Further, we can construct a scheme $\set S''$ for $\const M''=16$, $\beta''=20$, and $\const L''=\nicefrac{1}{16}$ as follows. Files are split  as $\vec X^{[16]} = ( (\vec X^{(1)}, \vec X^{(2)}), (\vec X^{(3)}, \vec X^{(4)}), \ldots, (\vec X^{(15)}, \vec X^{(16)}))$, and the scheme $\set S'$ runs on each of these pairs of files. If the user wants file $\vec X^{(4)}$, she keeps the answer corresponding to $(\vec X^{(3)}, \vec X^{(4)})$ and obtains $\hat{\vec X}^{(4)}$ with distortion $\const D''=\const D'$. The rate is $\const{R}''=8 \const{R}' = 3.60$ and the leakage is $\const{L}''=  \const{L}'\cdot\nicefrac{1}{8} = \nicefrac{1}{16}$.

\subsection{LP-Based Method From Lossy Compressors}
\label{sec:rates-from-lossy-compressors}



A lossy compressor that takes as input $\beta$ bits and compresses them to an encoded output that allows for reconstruction with per-bit distortion $D$ is in fact equivalent to a response function $A$ with distortion $D$. The rate of $A$ can be calculated as in \eqref{eq:response-function-rate-def}. Next, if we have a set of lossy compressors, each accepting an input of $\beta$ bits but allowing for different distortion levels, we are exactly in the situation considered  in \cref{sec:beyond-filtering}. Therefore, we can find schemes for arbitrary $\const M$ files and leakage level $\const L$ by solving the LP problem as described there.

\subsection{Examples of Lossy Compressors}
\label{sec:examples-lossy-compressors}

The method in \cref{sec:rates-from-lossy-compressors} only needs a lossy compressor(s). Below, we describe two ways to obtain them.

\subsubsection{Nonconstructive}\label{sec:KV-lossy-compressors}

In the context of finite-length information theory, the authors in \cite{KostinaVerdu12_1} 
derived an achievable rate of lossy compression for any finite block length $\beta$ and thus, proved existence of a corresponding lossy compressor. The special case of the source $\set{X}\sim\Uniform{\{0,1\}}$ is addressed in~\cite[Cor.~17]{KostinaVerdu12_1}. However, the fidelity criterion used in~\cite[Cor.~17]{KostinaVerdu12_1} is the excess-distortion probability, while most of the works in rate-distortion theory, as well as our work, focus on the average distortion.
Since for a nonnegative RV $Z$, it holds that $\eE{Z} = \int_0^\infty \ePr{Z > z} \dd z$, we can derive results for the average distortion criterion from the results in \cite{KostinaVerdu12_1}. 

We remark here that there do exist more advanced finite-length lossy compressors than the one from \cite[Cor.~17]{KostinaVerdu12_1}, see, e.g., \cite[Thm.~5]{MatsutaUyematsu15_1}. However, our focus here is on the construction of response functions for LWPIR schemes rather than on the best finite-length lossy compression scheme, which by itself is an interesting research direction in finite-length information theory.

\subsubsection{Simulated Annealing Based Search}\label{sec:simulate-annealing-search}

As an alternative, we implemented a search based on simulated annealing among a restricted set of deterministic functions (these functions can be seen as lossy compressors). More precisely, for a fixed number of input bits $\beta$ and rate $\const R$ (such that $\beta \const R$ is an integer), this set consists of functions mapping $\beta$-bit strings to $[2^{\beta \const R}]$ with the additional requirement that the pre-image set of every element in $[2^{\beta \const R}]$ has size $2^{\beta(1-\const R)}$. Such functions obviously have rate $\const R$ but different distortions when used as lossy compressors. 

The simulated annealing search works as follows. We start with a randomly generated function (satisfying the aforementioned requirements) and proceed with a random walk on the set of such functions. At each step, the next function to consider is constructed from the current one by randomly swapping two elements of different pre-images. If the new function has smaller distortion, it becomes the current function. Otherwise, it can be still accepted (with some probability that changes during the iterations). The iterations continue until a maximum number of iterations is reached. The search is restarted many times and the function with the smallest distortion is stored. The best functions for different values of $\const R$ then form the pool of lossy compressors.

\subsection{Numerical Results}

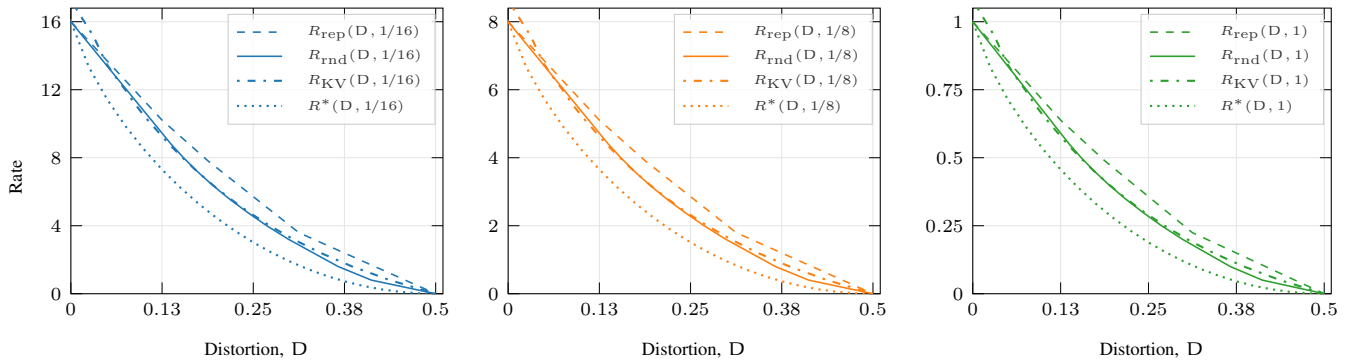
\begin{figure*}[t!]
	\subfloat%
	{%
		\input{\Figs/M16beta20L1o16.tikz}
	}
	\hfill
	\subfloat%
	{%
		\input{\Figs/M16beta20L1o8.tikz}
	}
	\hfill
	\subfloat%
	{%
		\input{\Figs/M16beta20L1.tikz}
	}
	\vspace{-2ex}
	\caption{Download rate versus distortion of three finite-size LWPIR schemes for $\const M=16$, $\beta =20$ bits, and leakage $\const L \in \{\nicefrac{1}{16},\nicefrac{1}{8},1\}$ (from left to right). For comparison, the corresponding asymptotic curves $R^\ast(\const{D},\const{L})$, obtained from \cref{thm:minimum-download_SSLWPIR} and \cref{cor:2-special-cases}, are also depicted.}
	\label{fig:M16beta20}
\end{figure*}

	In \cref{fig:M16beta20}, we plot achievable rate curves corresponding to different LWPIR schemes for $\const M=16$ files of $\beta=20$ bits each.
	
	First, similarly to the example in \cref{sec:LWPIR_from_smaller_ones}, we constructed other schemes 
	from the optimal schemes obtained with the LP method for $\const M \beta \le 4$ (cf.~Appendix~\ref{sec:exampl_small-optimal-schemes}). 
	The corresponding rate-distortion-leakage curves are labeled by $R_{\mathrm{rep}}(\const{D},\const{L})$. This approach is inferior to the two below (except for very small distortions where the next one produces worse rates). On the other hand, it provides easier constructions.
        
    Second, we used the WPIR+LC approach outlined in \cref{sec:WPIR-with-LC} 
    based on the lossy compressor from \cref{sec:KV-lossy-compressors}. The curves corresponding to the schemes obtained in this way are labeled by $R_{\textrm{KV}}(\const{D},\const{L})$. Note that this approach comes at the price of being nonconstructive, since the achievable rates presented in~\cite[Cor. 17]{KostinaVerdu12_1} are based on random coding arguments, while the approaches above and below give constructive schemes.
	
	Third, we used the lossy compressors found by the search described in \cref{sec:simulate-annealing-search} together with the LP-based method in \cref{sec:rates-from-lossy-compressors}. The corresponding curves are labeled by $R_{\textrm{rnd}}(\const D, \const L)$. This approach gives rates comparable to the previous one but is constructive, i.e., it produces particular schemes.

	Finally, we also plot the corresponding asymptotic curves $R^\ast(\const{D},\const{L})$ (i.e., when $\beta \to \infty$), obtained from \cref{cor:2-special-cases} (for $\const L=1$ and $\nicefrac{1}{16}$) or by numerically solving the optimization problem in \cref{thm:minimum-download_SSLWPIR} as described in \cref{sec:solving-opt-problem} with Gurobi \cite{gurobi} (for $\const L=\nicefrac 18$). The asymptotic curves serve as lower bounds on the finite-size curves.
	
	As a reminder, the curves for $\const L=1$ (``no privacy'') essentially represent different lossy compression schemes and serve as upper bounds for the finite-length rate-distortion function.

\section{Conclusion and Future Work}
\label{sec:conclusion}

We proposed to simultaneously relax the conditions of perfect retrievability and full privacy of standard PIR, referred to as LWPIR,  in order to obtain improved download rates for the case when all files are stored uncoded on a single server. In particular, we characterized the optimal rate-distortion-leakage tradeoff for an arbitrary number of asymptotically large files, and also proposed a method based on LP to construct schemes for a finite file size and an arbitrary number of files. By leveraging the LP approach, provably optimal LWPIR schemes and corresponding closed-form expressions for the rate-distortion-leakage tradeoff  were derived for the case when the database contains two or four bits.  Finally, for a database that contains $320$ bits, we compared  two construction methods based on the LP approach with a nonconstructive scheme downloading   subsets of files using a finite-length lossy compressor based on random coding.


The study of rate-distortion theory in the context of PIR is significant since it allows for further reducing the communication cost. One direct future work is to solve the remaining open problems in this paper, e.g., determining the exact closed-form expression for the optimal rate-distortion-leakage tradeoff when the file size is infinite. Also, when the leakage is a reciprocal of an integer, it is interesting to investigate the fundamental property for a distortion measure to assert the optimality of optimal lossy compression on top of a WPIR scheme. Moreover, several natural extensions of single-server LWPIR, including multiple replicated noncolluding and colluding servers, are of great interest for future research. On the other hand, as a by-product of the constructive optimal LWPIR schemes when the database contains at most four bits, we find that the proposed LP approach is of independent interest in the context of finite-length lossy compression. Here, the obtained optimal LWPIR scheme with $\const{L}=1$ serves as an optimal lossy source coding scheme that achieves the minimum average distortion value of a given rate and a fixed code length $\beta$. This is similar to the research line of~\cite{PolyanskiyPoorVerdu10_1,LinMoserChen18_1}, where the objective is to determine the best possible codes that minimize the average error probability for small-to-moderate block lengths of a given channel. Given a source coding rate and a finite code length, the source code structure that minimizes the average distortion value remains an open problem.


\appendices
\section{Uniform Distortion for Each File}
\label{sec:properties_LWPIR}

\begin{prop}[uniform distortion]
  \label{prop:uniform-Ds}
  Assume there is a scheme with distortion $D$. Then there exists a scheme $\set S'$ with uniform per-file distortion
  \[
    D^{(m)}_{\set S'} = D, \forall\,m\in [\const M],
  \]
  while the leakage and the rates are the same for both schemes.
\end{prop}
\begin{IEEEproof}
  For the sake of simplicity, we prove the proposition for $\const M=2$.
  
  Assume we have a scheme $\set S_1$ with $D^{(1)}_{\set S_1} \neq D^{(2)}_{\set S_1}$. The distortion of the scheme is
  \[
    D = D_{\set S_1} = \frac{D^{(1)}_{\set S_1} + D^{(2)}_{\set S_1}}{2}.
  \]
  
  We can construct a scheme $\set S_2$ where the handling of the files $\vec X^{(1)}$ and $\vec X^{(2)}$ is swapped. More precisely, 
  \begin{gather*}
    P^{(\set S_2)}_{Q|M=1} = P^{(\set S_1)}_{Q|M=2}, \quad
    P^{(\set S_2)}_{Q|M=2} = P^{(\set S_1)}_{Q|M=1}, \\
    A_{\set S_2}(Q, \vec X^{(1)}, \vec X^{(2)}) = A_{\set S_1}(Q, \vec X^{(2)}, \vec X^{(1)}), \\
    \hat{\vec X}^{(1)}_{\set S_2}(Q,A) = \hat{\vec X}^{(2)}_{\set S_1}(Q,A), \quad
    \hat{\vec X}^{(2)}_{\set S_2}(Q,A) = \hat{\vec X}^{(1)}_{\set S_1}(Q,A).
  \end{gather*}
  Note that the leakage and the rate of the scheme $\set S_2$ are the same as in $\set S_1$. However, the per-file distortions are swapped:
  \[
    D_{\set S_2}^{(1)} = D_{\set S_1}^{(2)} \text{ and } D_{\set S_1}^{(1)} = D_{\set S_2}^{(2)}.
  \]
  Now, we can apply time-sharing to  $\set S_1$ and $\set S_2$ with a time-sharing
coefficient of $\alpha = \nicefrac 12$. The per-file distortions of the resulting scheme $\set S'$ are both equal to $D$, being the arithmetic average of $D_{\set S_1}^{(1)}$ and $D_{\set S_1}^{(2)}$.
  
  The proof for $\const M > 2$ follows along the same lines but considers all schemes associated with all permutations of the files.
\end{IEEEproof}

\section{Proof of Theorem~\ref{thm:minimum-download_SSLWPIR}}
\label{sec:proof_thm1}

The achievability proof is clear from the sketch of proof. Here, we only need to prove the converse part and the upper bound on $\ecard{\set{Q}}$.

\subsubsection{Converse}
\label{sec:converse} 


To prove the converse part to~\cref{thm:minimum-download_SSLWPIR}, we first note that it was already shown in the converse proof of~\cite[Thm.~1]{WengYakimenkaLinRosnesKliewer20_1sub} (see~\cite[App.~B]{WengYakimenkaLinRosnesKliewer20_1sub}) that
\begin{IEEEeqnarray*}{c}
  R^\ast(\const{D},\const{L})\geq\min_{P_{Q|M},P_{\hat{X}^{[\const{M}]}\mid X^{[\const{M}]},Q}\in\set{F}(\const{D},\const{L})}\bigMIcond{X^{[\const{M}]}}{\hat{X}^{[\const{M}]}}{Q},\label{eq:RDL-constraints}
\end{IEEEeqnarray*}
for any $\beta$, where $X^{[\const{M}]}\sim\Uniform{\set{X}^{\const{M}}}$ and
\begin{IEEEeqnarray*}{rCl}
  \set{F}(\const{D},\const{L})& \eqdef &\Bigl\{\bigl(P_{Q|M}, P(\hat{x}^{[\const{M}]}\mid x^{[\const{M}]},q)\bigr)\colon\nonumber\\
  &&\quad\>\bigE[M,Q]{d(X^{(M)},\hat{X}^{(M)})}\leq\const{D},\,
  L(P_{Q|M})\leq\const{L}\Bigr\}.\label{eq:RDL-constraints}
\end{IEEEeqnarray*}


Next, consider an arbitrary $\bigl(P_{Q|M},P(\hat{x}^{[\const{M}]}\mid x^{[\const{M}]},q)\bigr)\in\set{F}(\const{D},\const{L})$. By definition, each distribution $P(\hat{x}^{[\const{M}]}\mid x^{[\const{M}]},q)$, $q\in\set{Q}$, results in the distortions
\begin{IEEEeqnarray*}{rCl}
  \tilde{\const{D}}^{(m)}_q& \eqdef &\sum_{x^{(m)},\hat{x}^{(m)}}P(x^{(m)}|q)P(\hat{x}^{(m)}|x^{(m)},q)d(x^{(m)},\hat{x}^{(m)}),\IEEEeqnarraynumspace
\end{IEEEeqnarray*}
for $m\in[\const{M}]$, satisfying the overall constraint
\begin{IEEEeqnarray*}{rCl}
  \sum_{m\in[\const{M}]}\sum_{q\in\set{Q}}P(m,q)\tilde{\const{D}}^{(m)}_q\leq \const{D}.
\end{IEEEeqnarray*}

Now, let $\constb{D}_q\eqdef\bigl(\const{D}^{(1)}_q,\ldots,\const{D}^{(\const{M})}_q\bigr)$, $q\in\set{Q}$, denote an $\const{M}$-dimensional vector of nonnegative distortion values. Define
\begin{IEEEeqnarray*}{rCl}
  \IEEEeqnarraymulticol{3}{l}{%
    \set{D}(\const{D},P_{Q|M})}\nonumber\\*\quad%
  & \eqdef &\Biggl\{\{\constb{D}_q\}_{q\in\set{Q}}\colon\sum_{q\in\set{Q}}\sum_{m\in[\const{M}]}P_{M,Q}(m,q)\const{D}^{(m)}_q\leq\const{D}\Biggr\}
\end{IEEEeqnarray*}
of such vectors that satisfy a given overall distortion constraint $\const{D}$ for a given conditional distribution $P_{Q|M}$. Let $\tilde{\constb{D}}_q =\bigl(\tilde{\const{D}}^{(1)}_q,\ldots,\tilde{\const{D}}^{(\const{M})}_q\bigr)$. It follows that

\begin{IEEEeqnarray}{rCl}
  \IEEEeqnarraymulticol{3}{l}{%
    \bigMIcond{X^{[\const{M}]}}{\hat{X}^{[\const{M}]}}{Q}}\nonumber\\*\quad%
  & = &\sum_{q\in\set{Q}}P_Q(q)\bigMIcond{X^{[\const{M}]}}{\hat{X}^{[\const{M}]}}{Q=q}
  \nonumber\\
  & \stackrel{(a)}{\geq} &\sum_{q\in\set{Q}}P_Q(q)\RDF_{X^{[\const{M}]}}(\tilde{\constb{D}}_q)
  \nonumber\\
  & \stackrel{(b)}{\geq} &\min_{\substack{\{\constb{D}_q\}_{q\in\set{Q}}\in\\*\set{D}(\const{D},P_{Q|M})}}\sum_{q\in\set{Q}}P_Q(q)\RDF_{X^{[\const{M}]}}(\constb{D}_q),\label{eq:MI_LB}
\end{IEEEeqnarray}
where $(a)$ follows from the definition of the joint rate-distortion function~\cite{Gray73_1}, and $(b)$ holds since $\{\tilde{\constb{D}}_q\}_{q\in\set{Q}}\in\set{D}(\const{D},P_{Q|M})$.

Therefore, taking the minimization on both sides of~\eqref{eq:MI_LB} with respect to $\bigl(P_{Q|M}, P(\hat{x}^{[\const{M}]}\mid x^{[\const{M}]},q)\bigr)\in\set{F}(\const{D},\const{L})$ yields
\begin{IEEEeqnarray*}{rCl}
  \IEEEeqnarraymulticol{3}{l}{%
    R^\ast(\const{D},\const{L})}\nonumber\\*%
  & \geq &\min_{\substack{P_{Q|M}\colon\\ L(P_{Q|M})\leq\const{L}}}\min_{\substack{\{\constb{D}_q\}_{q\in\set{Q}}\in\\*\set{D}(\const{D},P_{Q|M})}}\sum_{q\in\set{Q}}P_Q(q)\RDF_{X^{[\const{M}]}}(\constb{D}_q),
\end{IEEEeqnarray*}
for any $\beta$, where \eqref{eq:download-obj-ft} follows since $\RDF_{X^{[\const{M}]}}(\constb{D}_q)=\sum_{m\in[\const{M}]}\RDF_{X}(\const{D}_q^{(m)})$ as $\{X^{(m)}\}_{m\in[\const{M}]}$ are independent and identically distributed~\cite[Thm.~3.1]{Gray73_1}.

\subsubsection{Size Bound on Query Set}
\label{sec:size-bound_query}

We use the standard approach of applying Carath{\'e}odory's theorem~\cite[Thm.~15.3.5]{CoverThomas06_1} to limit the size of the query set $\set{Q}$ (these kind of bounding techniques are also discussed in~\cite[App.~C]{ElGamalKim11_1}). First, note that for all possible choices of query set $\set{Q}$, joint distribution $P_{M,Q}$, and distortion variables $\{\constb{D}_q\}_{q\in\set{Q}}$, it can easily be seen that the set of all achievable tuples $(\const{R},\const{D},\const{L})$ satisfying
\begin{IEEEeqnarray}{rCl}
  \const{R}& \geq &\sum_{q\in\set{Q}}P_Q(q)\RDF_{X^{[\const{M}]}}(\constb{D}_q),
  \nonumber\\
  \const{D}& \geq &\sum_{q\in\set{Q}}P_Q(q)\sum_{m\in[\const{M}]}P_{M|Q}(m|q)\const{D}_q^{(m)},
  \label{eq:RDL-expressions}\\
  \const{L}& \geq &\sum_{q\in\set{Q}}P_Q(q)\max_{m\in[\const{M}]}P_{M|Q}(m|q),\nonumber
\end{IEEEeqnarray}
is convex (again, see~\cite[App.~B]{WengYakimenkaLinRosnesKliewer20_1sub}). Now, consider a given choice for the query set $\set{Q}$, the joint distribution $P_{M,Q}=P_M P_{Q|M}$, and the distortion variables $\{\constb{D}_q\}_{q\in\set{Q}}$. To show that the right-hand side of \eqref{eq:RDL-expressions} can still be achieved with a certain restricted size of $\set{Q}$, we first define the $(\const{M}+2)$-dimensional 
 vector 
\begin{IEEEeqnarray*}{rCl}
  \vect{v}_q& \eqdef &\Bigl(\RDF_{X^{[\const{M}]}}(\constb{D}_q),\sum_{m\in[\const{M}]}P_{M|Q}(m|q)\const{D}_q^{(m)},\nonumber\\[1mm]
  &&\,\,\, \>\max_{m\in[\const{M}]}P_{M|Q}(m|q),P_{M|Q}(1|q),\ldots,P_{M|Q}(\const{M}-1|q)\Bigr)\IEEEeqnarraynumspace
\end{IEEEeqnarray*}
for each $q\in\set{Q}$. Note that the probabilities
\begin{IEEEeqnarray*}{c}
  P_M(m)=\sum_{q\in\set{Q}}P_Q(q)P_{M|Q}(m|q),\quad\forall\,m\in[\const{M}],
\end{IEEEeqnarray*}
are given. Then, let 
\begin{IEEEeqnarray*}{rCl}
  \vect{v}& = &\sum_{q\in\set{Q}}P_Q(q)\vect{v}_q
  \nonumber\\
  & = &\Biggl(\sum_{q\in\set{Q}}P_Q(q)\RDF_{X^{[\const{M}]}}(\constb{D}_q),\nonumber\\
  && \quad\>\sum_{q\in\set{Q}}P_Q(q)\sum_{m\in[\const{M}]}P_{M|Q}(m|q)\const{D}_q^{(m)},\nonumber\\
  && \quad\>\sum_{q\in\set{Q}}P_Q(q)\max_{m\in[\const{M}]}P_{M|Q}(m|q),\nonumber\\
  && \quad\>\sum_{q\in\set{Q}}P_Q(q)P_{M}(1|q),\ldots,\sum_{q\in\set{Q}}P_Q(q)P_{M}(\const{M}-1|q)\Biggr) \IEEEeqnarraynumspace
\end{IEEEeqnarray*}
be a convex combination of $\vect{v}_{q_1},\dotsc,\vect{v}_{q_{\ecard{\set{Q}}}}$.  
Now, Carath{\'e}odory's theorem states that any $d$-dimensional vector in a convex set $\set{V}$ can be represented as a convex combination of at most $d+1$ points in $\set{V}$. Hence, it follows that the size of $\set{Q}$ can be reduced to being at most $\const{M}+3$, while keeping $\vect{v}$ the same, i.e., the values of the right-hand side of \eqref{eq:RDL-expressions} and the given values of $P_M(m)$, $m\in[\const{M}]$, are unchanged. This completes the proof.

\section{Proof of \cref{cor:2-special-cases}}
\label{sec:proof-cor:2-special-case}
	
\subsubsection{``No-Leakage'' Case ($\const L = \nicefrac{1}{\const M}$)}

In this case, the leakage constraint (\ref{eq:download-L-constr}) reduces to
\[
  \sum_{q \in \set Q} \max_{m \in [\const M]} P_{Q|M}(q|m) \le 1.
\]

For any $q\in\set{Q}$, if we choose an arbitrary $m_0 \in [\const M]$, by definition we have $P_{Q|M}(q|m_0) \le \max_{m \in [\const M]} P_{Q|M}(q|m)$, which leads to
\[
  1 = \sum_{q \in \set Q} P_{Q|M} (q|m_0) \le \sum_{q \in \set Q} \max_{m \in [\const M]} P_{Q|M}(q|m) \le 1.
\]

This is only possible if and only if $P_{Q|M}(q|m_0) = \max_{m \in [\const M]} P_{Q|M}(q|m)$, for all $q \in \set Q$. Since $m_0$ is chosen arbitrarily, we can conclude that for all $q\in\set{Q}$, we have 
\begin{IEEEeqnarray*}{c}
  P_{Q|M}(q|m)= 
  \max_{m' \in [\const M]}P_{Q|M}(q|m') = P_Q(q),\quad\forall\,m\in[\const{M}].
\end{IEEEeqnarray*}

In other words,  RVs $Q$ and $M$ are independent and the leakage constraint (\ref{eq:download-L-constr}) is satisfied (with equality).  Thus, \eqref{eq:minimizing-download_SSLWPIR} becomes 
\begin{IEEEeqnarray}{rCl}
  \min_{P_{Q|M}} \min_{\{ \const{D}_q^{(m)} \}} & \quad & \sum_{q \in \set Q} \left( P_Q(q) \sum_{m \in [\const M]} \RDF_X(\const{D}_q^{(m)}) \right)
  \notag \\
  \text{s.t.}& & \sum_{q \in \set Q} \sum_{m \in [\const M]} \frac{P_{Q}(q)}{\const M} \const{D}_q^{(m)} \le \const{D}. \label{eq:Dconstraint}
\end{IEEEeqnarray}

Since $\sum_{q \in \set Q} P_Q(q) = 1$, it follows that 
\[
  \sum_{q \in \set Q} \sum_{m \in [\const M]} \frac{P_{Q}(q)}{\const M} = 1.
\]
Moreover,  the objective function is bounded from below as 
\begin{IEEEeqnarray}{rCl}
  \IEEEeqnarraymulticol{3}{l}{%
    \sum_{q \in \set Q} \left( P_Q(q) \sum_{m \in [\const M]} \RDF_X(\const{D}_q^{(m)}) \right) }\nonumber\\*\quad%
  & = &\const M \sum_{q \in \set Q} \sum_{m \in [\const M]} \frac{P_Q(q)}{\const M} \RDF_X(\const{D}_q^{(m)})
  \nonumber\\
  & \stackrel{(a)}{\ge} &\const M \, \RDF_X \left( \sum_{q \in \set Q} \sum_{m \in [\const M]}\frac{P_Q(q)}{\const{M}}\const{D}_q^{(m)} \right) 
  \stackrel{(b)}{\geq}  \const M \, \RDF_X(\const D),
  \IEEEeqnarraynumspace\label{eq:LB_M-RD_X}
\end{IEEEeqnarray}
where $(a)$ follows from the convexity of $\RDF_X$ and Jensen's inequality~\cite[Sec.~2.6]{CoverThomas06_1}; and $(b)$ holds because of \eqref{eq:Dconstraint} and the fact that $\RDF_X$ is nonincreasing.  
Finally, the lower bound in~\eqref{eq:LB_M-RD_X} is achievable by choosing 
$\const{D}_q^{(m)} = \const D$ (and any distribution of $Q$), which concludes the proof for the ``no-leakage'' case.

	
\subsubsection{``No-Privacy'' Case ($\const L = 1$)}
	

Here, we directly show that the objective function in \eqref{eq:download-obj-ft} can be bounded from below by
\begin{IEEEeqnarray*}{rCl}
  \IEEEeqnarraymulticol{3}{l}{%
    \sum_{q\in\set Q} \left( P_Q(q)\sum_{m \in [\const M]} \RDF_X(\const{D}_q^{(m)}) \right)
  }\nonumber\\*\quad%
  & \overset{(a)}{=} &\sum_{q \in \set Q} \left( \Biggl( \sum_{m' \in [\const M]} \frac{P_{Q|M}(q|m')}{\const M} \Biggr) \sum_{m \in [\const M]} \RDF_X(\const{D}_q^{(m)}) \right)
  \nonumber\\
  & \overset{(b)}{\ge} &\sum_{q \in \set Q} \sum_{m \in [\const M]} \frac{P_{Q|M}(q|m)}{\const M} \RDF_X(\const{D}_q^{(m)})
  \nonumber\\
  & \overset{(c)}{\ge} &\RDF_X \Biggl( \sum_{q \in \set Q} \sum_{m \in [\const M]} \frac{P_{Q|M}(q|m)}{\const M} \const{D}_q^{(m)} \Biggr) \overset{(d)}{\ge} \RDF_X(\const D).
  \IEEEeqnarraynumspace\label{eq:LB_RD_X}
\end{IEEEeqnarray*}
Here, $(a)$ follows from the law of total probability; $(b)$ is the result of dropping all other terms in the sum of $\sum_{m' \in [\const M]} \nicefrac{P_{Q|M}(q|m')}{\const M}$ except for $\nicefrac{P_{Q|M}(q|m)}{\const M}$; $(c)$ is from the convexity of $\RDF_X$ and Jensen's inequality~\cite[Sec.~2.6]{CoverThomas06_1}; and $(d)$ is from~\eqref{eq:download-D-constr} and the fact that $\RDF_X$ is nonincreasing.
	
Finally, the value $\RDF_X(\const D)$ of the objective function is attained by setting $P_{Q|M}(q|m_0) = 1$  for exactly one $m_0\in[\const{M}]$ (and zero otherwise) for any $q\in\set{Q}$, and $\const D_q^{(m)}\eqdef\const D$, which concludes the proof for the ``no-privacy'' case.

\section{Proof of \cref{lem:vertices-set_query-set}} \label{sec:proof-lemma-1}
In the following, for notational convenience, let $p_{qm} \eqdef P(q|m)$.
  Fix some feasible solution $\{p_{qm}\}$, $\{\xi_q\}$ of \eqref{eq:LP-formulation} and consider $q_0 \in \set Q \setminus \set{Q}_\textnormal{v}$, i.e., $\vec c_{q_0}$ is not a vertex of $\set P(\set V)$. Then, $\vec c_{q_0}$ can be expressed as a convex combination of vertices (the coefficients are nonnegative and sum up to $1$) as 
  \[
    \vec c_{q_0} = \sum_{q \in \set{Q}_\textnormal{v}} \alpha_q \vec c_q + \alpha_\mathrm{max} \vec c_\mathrm{max}.
  \]
  From this, for all $m \in [\const M]$,
  \begin{IEEEeqnarray*}{rCl}
    D_{q_0}^{(m)} &=& \sum_{q \in \set{Q}_\textnormal{v}} \alpha_q D_q^{(m)} + \alpha_\mathrm{max} \max_{q \in \set Q} D_q^{(m)} \ge \sum_{q \in\set{Q}_\textnormal{v}} \alpha'_q D_q^{(m)}, \\
    R_{q_0} & \ge & \sum_{q \in\set{Q}_\textnormal{v}} \alpha'_q R_q, \quad
    \alpha'_q \eqdef \frac{\alpha_q}{\sum_{q' \in \set{Q}_\textnormal{v}}\alpha_{q'}},\quad q \in \set{Q}_\textnormal{v}.
  \end{IEEEeqnarray*}

  Now, construct a new solution $\{p'_{qm}\}$, $\{\xi'_q\}$ by redistributing probability mass from $p_{q_0m}$ as follows:
  \begin{IEEEeqnarray*}{rCl}
    p'_{qm} &=& \begin{cases}
      p_{qm} + \alpha'_q p_{q_0m} & \text{ if } q \in \set{Q}_\textnormal{v},\\
      0 & \text{ if } q = q_0,\\
      p_{qm} & \text{ otherwise},
    \end{cases}
    \\[1mm]
    \xi'_{q} &=& \begin{cases}
      \xi_{q} + \alpha'_q \xi_{q_0} & \text{ if } q \in \set{Q}_\textnormal{v},\\
      0 & \text{ if } q = q_0,\\
      \xi_{q} & \text{ otherwise}.
    \end{cases}
  \end{IEEEeqnarray*}
  It can be straightforwardly verified that this is also a feasible solution and that the corresponding objective value is not larger than the original one. Therefore, we can remove all variables $p_{q_0m}$, $\xi_{q_0}$ from \eqref{eq:LP-formulation}. In the same manner, we can remove all the variables corresponding to all $q \in \set Q \setminus \set{Q}_\textnormal{v}$ and further assume $\set Q = \set Q_\textnormal{v}$. Note that there is no query corresponding to $\vec c_{\textrm{max}}$ in the LP.

\section{Proof of \cref{thm:opt-tradeoff-M2-beta1}} \label{app:M2_beta1}

Before proving the theorem, we recall a well-known result about LP. Consider the conventional LP problem
\begin{subequations}
  \label{eq:standard-LP}
  \begin{IEEEeqnarray}{rCl}
    \min_{\vec x} & &\quad \vect{c}^\intercal \vec x\\
    \textnormal{s.t.} & &\quad \mat{A} \vec x = \vec b, \\
    & &\quad\vec x \ge \bm 0,
  \end{IEEEeqnarray}
\end{subequations}%
where matrix $\mat{A} \in \Reals^{k \times n}$, $\rank{\mat{A}} = k \le n$, $\vec x \in \Reals^n$, and $\vec b \in \Reals^k$. 

Let $\set B\subseteq [n]$ be some set of $k$ linearly independent columns in $\mat{A}$ and define $\set{N}\eqdef [n] \setminus \set{B}$. With sets $\set{B}$ and $\set{N}$, without loss of generality, \eqref{eq:standard-LP} can be written  as
\begin{IEEEeqnarray*}{rCl}
  \min_{\vec x} & &\quad \trans{\vect{c}}_{\set{B}}\vect{x}_\set{B}+\trans{\vect{c}}_{\set{N}}\vect{x}_\set{N}
  \\
  \textnormal{s.t.} & &\quad \mat{A}_{\set{B}}\vect{x}_{\set{B}}+\mat{A}_{\set{N}}\vect{x}_\set{N} = \vec b, \\
  & &\quad\vect{x}_{\set{B}},\vect{x}_{\set{N}} \ge \bm 0,
\end{IEEEeqnarray*}
where  $\mat{A}_\set{J}$ represents the submatrix of $\mat{A}$ restricted in columns by the set $\set{J}$, and $\vect{x}_{\set{J}}$ is the subvector of $\vect{x}$ with entries indexed by $\set J$.

\begin{prop}[{\cite[p.~44]{LuenbergerYe16_1}}]
  \label{prop:lp-optimality}	
  If all the entries of the vectors $\mat{A}_{\set B}^{-1} \vec b$ and $\vec c_{\set N}^\intercal - \vec c_{\set B}^\intercal \mat{A}_{\set B}^{-1}\mat{A}_{\set N}$ are nonnegative, 
  then the optimal solution to the LP is 
    $\vec c_{\set B}^\intercal \mat{A}_{\set B}^{-1} \vec b$.
\end{prop}


\begin{table}[t!]
  \caption{All set partition functions of $\{ 0,1 \}^{2}$. The functions with no tabulated value for  $q$  are not used in the LP formulation.}
  \label{tab:15-partitions}
  \vspace{-2ex}
  \centering
  
  \begin{tabular}{lllll}
    \toprule
    $q$ & subsets of a partition & $R_q$ & $D_q^{(1)}$ & $D_q^{(2)}$\\
    \midrule
    $q_1$ & $\{00\},\{01\},\{10\},\{11\}$ & 2 & 0 & 0
    \\
    \\
        & $\{00,11\},\{01\},\{10\}$
                                 & $\nicefrac 32$ & $\nicefrac 14$ & $\nicefrac 14$ \\
        & $\{00,10\},\{01\},\{11\}$
                                 & $\nicefrac 32$ & $\nicefrac 14$ & 0 \\
        & $\{00,01\},\{10\},\{11\}$
                                 & $\nicefrac 32$ & 0              & $\nicefrac 14$ \\
        & $\{00\},\{01,11\},\{10\}$
                                 & $\nicefrac 32$ & $\nicefrac 14$ & 0 \\
        & $\{00\},\{01,10\},\{11\}$
                                 & $\nicefrac 32$ & $\nicefrac 14$ & $\nicefrac 14$ \\
        & $\{00\},\{01\},\{10,11\}$
                                 & $\nicefrac 32$ & 0              & $\nicefrac 14$ \\
    \\
        & $\{00,01,11\},\{10\}$
                                 & 1 & $\nicefrac 14$ & $\nicefrac 14$ \\
        & $\{00,01,10\},\{11\}$
                                 & 1 & $\nicefrac 14$ & $\nicefrac 14$ \\
        & $\{00,10,11\},\{01\}$
                                 & 1 & $\nicefrac 14$ & $\nicefrac 14$  \\
        & $\{00\},\{01,10,11\}$ & 1 & $\nicefrac 14$ & $\nicefrac 14$ \\
    \\
    $q_2$ & $\{00,01\},\{10,11\}$
                                 & 1 & 0              & $\nicefrac 12$ \\
    $q_3$ & $\{00,10\},\{01,11\}$
                                 & 1 & $\nicefrac 12$ & 0 \\
        & $\{00,11\},\{01,10\}$
                                 & 1 & $\nicefrac 12$ & $\nicefrac 12$ \\
    \\
    $q_4$ & $\{00,01,10,11\}$
                                 & 0 & $\nicefrac 12$ & $\nicefrac 12$ \\
    \bottomrule
  \end{tabular}
\end{table}

Now, we prove the theorem by solving the LP in \eqref{eq:LP-formulation}. Note that the total number of all possible set partitions of $\{0,1\}^{2}$ is $15$. We present all of them in \cref{tab:15-partitions}. Therein, we list the subsets of all the partitions. After filtering (by constructing the convex hull $\set{P}(\set{V})$ as outlined in Section~\ref{sec:fliter-function}), one can conclude that the vertex set of the convex hull $\set{P}(\set{V})$ is $\{ \vec c_{q_1}, \vec c_{q_2}, \vec c_{q_3}, \vec c_{q_4}\} \cup \{ (2, \nicefrac 12, \nicefrac 12)\}$. Thus, according to Lemma \ref{lem:vertices-set_query-set}, the query set $\set{Q}$ in~\eqref{eq:LP-formulation} can be set to $\set Q_{\textnormal v} = \{ q_1, q_2, q_3, q_4 \}$. 
For example, the convex representation of the vector corresponding to the partition $\{00,11\},\{01\},\{10\}$ is $\left( \nicefrac 32, \nicefrac 14, \nicefrac 14 \right) = \nicefrac 12 \left( 2, 0, 0 \right)+ \nicefrac 14 \left( 0, \nicefrac 12, \nicefrac 12 \right) + \nicefrac 14 (2, \nicefrac 12, \nicefrac 12)$. Similar expressions can be easily written for other functions that have been filtered out.

Next, we reformulate our problem in the form of \cref{eq:standard-LP} as follows:
\begin{subequations}
	\label{eq:LP-formulation-SF}
	\begin{IEEEeqnarray}{rCll}
		\min_{p_{qm}} & \quad & \frac{1}{\const M} \sum_{m \in [\const M]} \sum_{q \in \set Q} p_{qm} R_q
		\\
		\text{s.t.} &  & \sum_{q \in \set Q} p_{qm} = 1, \quad m \in [\const M], \label{eq:LP-SF-pmfs}
		\\
		& & \sum_{q \in \set Q} \xi_q = \const M \const L, \label{eq:LP-SF-constrL}\\
		& & \sum_{m \in [\const M]} \sum_{q \in \set Q} p_{qm} D_q^{(m)} = \const M \const{D}, \label{eq:LP-SF-constrD} \\
		& & p_{qm} - \xi_q + s_{qm} = 0, \quad q \in \set Q,\, m \in [\const M], \label{eq:LP-SF-p-xi-s} \\
		& & p_{qm},\, \xi_q,\, s_{qm} \ge 0,
	\end{IEEEeqnarray}
\end{subequations}
where we have used the short-hand notation $p_{qm} \eqdef P(q|m)$ for notational convenience.
Note that we have introduced slack variables $s_{qm}$  in \eqref{eq:LP-SF-p-xi-s}, which allows \eqref{eq:LP-SF-constrL} to become an equality.  Also, the inequality constraint in \eqref{eq:LP-formulation-D-constraint}  has been replaced by an equality constraint in (\ref{eq:LP-SF-constrD}). The latter is true if we assume $\const D \le \nicefrac 12$ and because $D_{q_4}^{(1)} = D_{q_4}^{(2)} = \nicefrac 12$, $R_{q_4}=0$. Indeed, we can always move the probability mass to $p_{q_4m}$, $m \in [\const M]$, until we reach equality to $\const D$, and the objective function will only decrease. We also assume $\const L \in [\nicefrac 1{\const M}, 1] = [\nicefrac 12, 1]$. The matrix $\mat A$ corresponding to this LP is presented in \cref{eq:proof-mtrx-A}. For visual clarity, we have substituted zeros with dots.

\NiceMatrixOptions{cell-space-top-limit = 2pt}
\NiceMatrixOptions{cell-space-bottom-limit = 2pt}
\begin{equation}
	\label{eq:proof-mtrx-A}
	\begin{pNiceArray}{cccccccc|cccc|cccccccc}[nullify-dots,small,first-row,last-col]
	    \Block{1-4}{p_{q1}} & & & & \Block{1-4}{p_{q2}} & & & & \Block{1-4}{\xi_q} & & & & \Block{1-8}{s_{qm}} \\
		1 & 1 & 1 & 1 & . & . & . & . & . & . & . & . & . & . & . & . & . & . & . & . & (\ref{eq:LP-SF-pmfs}) \\
		. & . & . & . & 1 & 1 & 1 & 1 & . & . & . & . & . & . & . & . & . & . & . & . & (\ref{eq:LP-SF-pmfs}) \\
		. & . & . & . & . & . & . & . & 1 & 1 & 1 & 1 & . & . & . & . & . & . & . & . & (\ref{eq:LP-SF-constrL})\\
		. & . & \frac{1}{2} & \frac{1}{2} & . & \frac{1}{2} & . & \frac{1}{2} & . & . & . & . & . & . & . & . & . & . & . & . & (\ref{eq:LP-SF-constrD}) \\
		\hline
		1 & . & . & . & . & . & . & . & -1 & . & . & . & 1 & . & . & . & . & . & . & . & (\ref{eq:LP-SF-p-xi-s}) \\
		. & 1 & . & . & . & . & . & . & . & -1 & . & . & . & 1 & . & . & . & . & . & . & (\ref{eq:LP-SF-p-xi-s}) \\
		. & . & 1 & . & . & . & . & . & . & . & -1 & . & . & . & 1 & . & . & . & . & . & \Vdots \\
		. & . & . & 1 & . & . & . & . & . & . & . & -1 & . & . & . & 1 & . & . & . & . \\
		. & . & . & . & 1 & . & . & . & -1 & . & . & . & . & . & . & . & 1 & . & . & . \\
		. & . & . & . & . & 1 & . & . & . & -1 & . & . & . & . & . & . & . & 1 & . & . \\
		. & . & . & . & . & . & 1 & . & . & . & -1 & . & . & . & . & . & . & . & 1 & . \\
		. & . & . & . & . & . & . & 1 & . & . & . & -1 & . & . & . & . & . & . & . & 1 & (\ref{eq:LP-SF-p-xi-s}) \\
	\end{pNiceArray}
\end{equation}
\vspace{2ex}

Consider the following choices of $\set B$.

\begin{itemize}
\item $\set B = \{1, 2, 4, 5, 7, 8, 9, 10, 11, 12, 15, 18\}$. Then, $\mat A_{\set B}^{-1} \vec b = 2(1-\const D-\const L,\const L-\nicefrac{1}{2},\const D,1-\const D-\const L,\const L-\nicefrac{1}{2},\const D,1-\const D-\const L,\const L-\nicefrac{1}{2},\const L-\nicefrac{1}{2},\const D,\const L-\nicefrac{1}{2},\const L-\nicefrac{1}{2}) \ge \bm 0$, provided $\const D \le 1-\const L$. Also $\vec c_{\set N}^\intercal - \vec c_{\set B}^\intercal \mat{A}_{\set B}^{-1}\mat{A}_{\set N} = \nicefrac 12 (0,0,1,2,1,1,2,1) \ge \bm 0$. Therefore, the solution $\vec c_{\set B}^\intercal \mat{A}_{\set B}^{-1} \vec b = 3-2\const L-4\const D$ is optimal.
  
\item $\set B = \{2,4,7,8,9,10,11,12,15,16,17,18\}$. In this case, $\mat A_{\set B}^{-1} \vec b = 2(\const L-\nicefrac{1}{2},1-\const L,\nicefrac 32-2\const D-\const L,2\const D+\const L-1,0,\const L-\nicefrac 12,\nicefrac 32 -2\const D-\const L,2 \const D+\const L-1,\nicefrac 32-2\const D-\const L,2\const D+2\const L-2,0,\const L-\nicefrac 12) \ge \bm 0$ if $1-\const L \le \const D \le \nicefrac 12$. Also $\vec c_{\set N}^\intercal - \vec c_{\set B}^\intercal \mat{A}_{\set B}^{-1}\mat{A}_{\set N} = \nicefrac 12 (1,1,1,1,0,0,0,0) \ge \bm 0$ and the solution $\vec c_{\set B}^\intercal \mat{A}_{\set B}^{-1} \vec b = 1-2\const D$ is optimal.
\end{itemize}

\section{Small Optimal Lossy Compressors}
\label{sec:exampl_small-optimal-schemes}

For small sizes of up to $4$ bits, we were able to exhaustively search for the best lossy compressors with fixed-size input and arbitrary-size output. In particular, we found the following compressors for $4$ bits of input. We present them by their Huffman codes.
\subsubsection{Compressor 1}
Any input from the set $\{0000, 0001, 1001, 0101, 0011\}$ is encoded as $\mathtt{00}$ and decoded back to $0001$ (giving an average per-bit distortion of $\nicefrac{1}{5}$ on these inputs); any input from the set $\{1110,1101,1011,0111,1111\}$ is encoded as $\mathtt{01}$ and decoded back to $1111$ (again, per-bit distortion of $\nicefrac{1}{5}$). The remaining inputs are encoded as follows: $1000$  as $\mathtt{110}$, $0100$ as $\mathtt{111}$, $1100$ as $\mathtt{1000}$, $0010$ as $\mathtt{1001}$, $1010$ as $\mathtt{1010}$, and $0110$ as $\mathtt{1011}$, and they are decoded back to themselves (with no distortion). With this encoding scheme, the length of the codeword is $2$, $3$, and $4$ with probabilities $\nicefrac{10}{16}$, $\nicefrac{2}{16}$, and $\nicefrac{4}{16}$, respectively, and the average per-bit codeword length (i.e., the rate) is $\const R = \nicefrac{21}{32}$.
The distortion is $\nicefrac{1}{5}$ with probability $\nicefrac{10}{16}$ and $0$ with probability $\nicefrac{6}{16}$ thus giving an average distortion of $\const D = \nicefrac 18$.

\subsubsection{Compressor 2}
Similarly, the encoding/decoding is as follows:
\begin{itemize}
	\item \{0000, 0010, 1010, 0110, 0011\} $\to \mathtt{00} \to 0010$,
	\item \{1000, 0001, 1001, 1101, 1011\} $\to \mathtt{01} \to 1001$,
	\item \{0100, 1100, 1110\} $\to \mathtt{10} \to 1100$,
	\item \{0101, 0111, 1111\} $\to \mathtt{11} \to 0111$.
\end{itemize}
The compressor has rate $\const R=\nicefrac 12$ and distortion $\const D=\nicefrac{3}{16}$.

\subsubsection{Compressor 3}
The encoding/decoding is as follows:
\begin{itemize}
	\item $\{ 1100,1010,0110,1110,1111 \} \to \mathtt 0  \to 1110$,
	\item \{0000, 1000, 0100, 0010, 0001, 1001, 0101, 1101, 0011, 1011, 0111\} $\to \mathtt 1 \to 0001$.
\end{itemize}
The compressor has rate $\const R=\nicefrac 14$ and distortion $\const D=\nicefrac{5}{16}$.

\subsubsection{No-Distortion Compressor}
The compressor encodes the input to itself, thus providing distortion $\const D=0$ and rate $\const R=1$.

\subsubsection{Random-Guess Compressor}
This is a virtual compressor, which encodes any input to an empty string, thus having $\const R=0$ and distortion of random guess $\const D=\nicefrac 12$.


In fact, with these five compressors used as response functions, we can construct optimal LWPIR schemes for the set-ups:
\begin{itemize}
	\item $\const M=4$ files of $\beta=1$ bit each,
	\item $\const M=2$ files of $\beta=2$ bits each,
	\item $\const M=1$ file of $\beta=4$ bits,
\end{itemize}
i.e., the set-ups with $\const M \beta = 4$.  Note only that the rates should be normalized by $\beta$.


\balance

\bibliographystyle{IEEEtran}
\bibliography{defshort1,biblioHY}

%
%
%
%
%
%
%

\end{document}

%% file: imports-and-defs_final.tex
\usepackage[utf8]{inputenc} 
\usepackage[T1]{fontenc}
\usepackage{url}
\usepackage{ifthen}
\usepackage[noadjust]{cite} 
\usepackage[cmex10]{amsmath} 
\usepackage{amsfonts,amssymb}

\usepackage{mathtools}

\usepackage{graphicx}
\usepackage{color,url}

\usepackage{datetime}

\usepackage{nicefrac}

\usepackage{enumitem} 

\usepackage{verbatim} 

\usepackage[mathscr]{euscript}

\usepackage{blkarray,bigstrut} 

\usepackage[algoruled,vlined,linesnumbered,nofillcomment]{algorithm2e} 

\usepackage[capitalize]{cleveref} 
	\crefname{equation}{}{}
	\crefname{theorem}{Theorem}{Theorems}
	\crefname{lemma}{Lemma}{Lemmas}
	\crefname{cor}{Corollary}{Corollaries}
	\crefname{prop}{Proposition}{Propositions}
	\crefname{note}{Note}{Notes}
	\crefname{appsec}{Appendix}{Appendices}
	\crefname{definition}{Definition}{Definitions}
	\crefname{conj}{Conjecture}{Conjectures}
	\crefname{construction}{Construction}{Constructions}

\usepackage{booktabs}  
\usepackage{tabularx}  
\usepackage{multirow}

\usepackage{bm}
\usepackage{bbm}
\usepackage[normalem]{ulem} 

\usepackage{balance}

\usepackage[caption=false,font=footnotesize]{subfig}

\usepackage[dvipsnames]{xcolor} 

\usepackage{tikz}
\usetikzlibrary{shapes,arrows}
\usetikzlibrary{matrix}
\usetikzlibrary{spy} 
\usepackage{pgfplots}
\usepgfplotslibrary{fillbetween}
\usepackage[draft]{tikzpeople} 
\usetikzlibrary{calc} 
\usetikzlibrary{positioning}
\usetikzlibrary{arrows}
\usetikzlibrary{shapes.multipart}
\usepackage{nicematrix} 

\allowdisplaybreaks

	\newtheorem{theorem}{Theorem}
	\newtheorem{lemma}{Lemma}
	\newtheorem{cor}{Corollary}
	\newtheorem{prop}{Proposition}

	\newtheorem{definition}{Definition}

\allowdisplaybreaks

\newcommand{\tn}[1]{\textnormal{#1}}        

\renewcommand{\vec}[1]{\bm{#1}}
\newcommand{\vect}[1]{\bm{#1}} 
\newcommand{\mat}[1]{\mathsf{#1}} 
\newcommand{\const}[1]{\textnormal{\usefont{U}{eur}{m}{n}\selectfont #1}} 
\newcommand{\constb}[1]{\textnormal{\usefont{U}{eur}{b}{n}\selectfont #1}} 

\newcommand{\set}[1]{\mathcal{#1}}           
\newcommand{\card}[1]{\left|#1\right|}       
\newcommand{\ecard}[1]{|#1|}

\newcommand{\EE}[2][]{\mathbb E_{#1} \left[ #2 \right]}

\newcommand{\eqdef}{\triangleq} 
\newcommand{\unif}[1]{\mathscr{U} \left( #1 \right)}

\newcommand{\dd}{\mathop{}\!\mathrm{d}} 
\newcommand{\trans}[1]{#1^{\intercal}} 
\newcommand{\rank}[1]{\mathrm{rank}\left(#1\right)} 
\DeclareRobustCommand{\cumBinom}{\genfrac<>{0pt}{}}

\DeclarePairedDelimiter\floor{\lfloor}{\rfloor}

\renewcommand{\Pr}[1]{\mathbb{P}\left[ #1 \right]}
\newcommand{\ePr}[1]{\mathbb{P}[#1]}

\newcommand{\BigPr}[1]{\mathbb{P}\Bigl[#1\Bigr]}

\newcommand{\Prcond}[2]{\mathbb{P}\left[#1 \kern0.1em\middle|\kern0.1em #2\right]}
\newcommand{\ePrcond}[2]{\mathbb{P}[#1 \kern0.1em|\kern0.1em #2]} 
\newcommand{\bigPrcond}[2]{\mathbb{P}\bigl[#1 \kern-0.1em \bigm| \kern-0.1em#2\bigr]}
\newcommand{\BigPrcond}[2]{\mathbb{P}\Bigl[#1 \kern-0.1em \Bigm| \kern-0.1em#2\Bigr]}
\newcommand{\biggPrcond}[2]{\mathbb{P}\biggl[#1 \kern-0.1em \biggm| \kern-0.1em#2\biggr]}
\newcommand{\BiggPrcond}[2]{\mathbb{P}\Biggl[#1 \kern-0.1em \Biggm| \kern-0.1em#2\Biggr]}

\newcommand{\Uniform}[1]{\mathscr{U}\left(#1\right)} 
\newcommand{\eUniform}[1]{\mathscr{U}(#1)} 
\newcommand{\bigUniform}[1]{\mathscr{U}\bigl(#1\bigr)}

\newcommand{\Exp}{\operatorname{\mathbb{{E}}}}

\newcommand{\E}[2][]{\Exp_{#1}\left[#2\right]}
\newcommand{\eE}[2][]{\Exp_{#1}[#2]}
\newcommand{\bigE}[2][]{\Exp_{#1}\bigl[#2\bigr]}
\newcommand{\BigE}[2][]{\Exp_{#1}\Bigl[#2\Bigr]}

\newcommand{\Econd}[3][]{\Exp_{#1}\left[#2 \kern0.1em\middle|\kern0.1em #3\right]}
\newcommand{\eEcond}[3][]{\Exp_{#1}[#2 \kern0.1em|\kern0.1em #3]}
\newcommand{\bigEcond}[3][]{\Exp_{#1}\bigl[#2 \kern-0.1em \bigm| \kern-0.1em #3\bigr]}
\newcommand{\BigEcond}[3][]{\Exp_{#1}\Bigl[#2 \kern-0.1em \Bigm| \kern-0.1em #3\Bigr]}
\newcommand{\biggEcond}[3][]{\Exp_{#1}\biggl[#2 \kern-0.1em \biggm| \kern-0.1em #3\biggr]}
\newcommand{\BiggEcond}[3][]{\Exp_{#1}\Biggl[#2 \kern-0.1em \Biggm| \kern-0.1em #3\Biggr]}

\newcommand{\HH}{\mathop{}\!\mathsf{H}} 
\newcommand{\Hb}{\HH_{\tn{b}}} 

\newcommand{\HPcond}[2]{\HH\left(#1 \kern0.1em\middle|\kern0.1em #2\right)}
\newcommand{\eHPcond}[2]{\HH(#1 \kern0.1em|\kern0.1em #2)} 
\newcommand{\bigHPcond}[2]{\HH\bigl(#1 \kern-0.1em \bigm| \kern-0.1em#2\bigr)}
\newcommand{\BigHPcond}[2]{\HH\Bigl(#1 \kern-0.1em \Bigm| \kern-0.1em#2\Bigr)}

\newcommand{\II}{\mathop{}\!\mathsf{I}}  
\newcommand{\MI}[2]{\II\left(#1 \kern0.1em{;}\kern0.1em #2\right)} 
\newcommand{\eMI}[2]{\II(#1 \kern0.1em{;}\kern0.1em #2)} 
\newcommand{\bigMI}[2]{\II\bigl(#1 \kern0.1em{;}\kern0.1em #2\bigr)}
\newcommand{\BigMI}[2]{\II\Bigl(#1 \kern0.1em{;}\kern0.1em #2\Bigr)}
\newcommand{\MIcond}[3]{\II\left(#1 \kern0.1em{;}\kern0.1em #2 \kern0.1em\middle|\kern0.1em #3\right)}
\newcommand{\eMIcond}[3]{\II(#1 \kern0.1em{;}\kern0.1em #2 \kern0.1em|\kern0.1em #3)} 
\newcommand{\bigMIcond}[3]{\II\bigl(#1 \kern0.1em{;}\kern0.1em #2 \kern-0.1em \bigm| \kern-0.1em#3\bigr)}
\newcommand{\BigMIcond}[3]{\II\Bigl(#1 \kern0.1em{;}\kern0.1em #2 \kern-0.1em \Bigm| \kern-0.1em#3\Bigr)}

\newcommand{\conv}[2]{\operatorname{conv}\left(#1 \kern0.1em{,}\kern0.1em #2\right)}

\newcommand{\LL}[2]{\EE[#1]{\ell^\ast\left(#2\right)}}

\DeclareMathOperator*{\argmax}{arg\,max}  

\DeclareMathOperator{\dist}{d} 

\newcommand{\Reals}{\mathbb{R}} 

\newcommand{\Rop}{R_{\mathrm{op}}}
\newcommand{\Rinf}{R_{\mathrm{inf}}}
\newcommand{\RDF}{\mathfrak{R}}

\makeatletter
\tikzset{
	database/.style={
		path picture={
			\draw (0, 1.5*\database@segmentheight) circle [x radius=\database@radius,y radius=\database@aspectratio*\database@radius];
			\draw (-\database@radius, 0.5*\database@segmentheight) arc [start angle=180,end angle=360,x radius=\database@radius, y radius=\database@aspectratio*\database@radius];
			\draw (-\database@radius,-0.5*\database@segmentheight) arc [start angle=180,end angle=360,x radius=\database@radius, y radius=\database@aspectratio*\database@radius];
			\draw (-\database@radius,1.5*\database@segmentheight) -- ++(0,-3*\database@segmentheight) arc [start angle=180,end angle=360,x radius=\database@radius, y radius=\database@aspectratio*\database@radius] -- ++(0,3*\database@segmentheight);
		},
		minimum width=2*\database@radius + \pgflinewidth,
		minimum height=3*\database@segmentheight + 2*\database@aspectratio*\database@radius + \pgflinewidth,
	},
	database segment height/.store in=\database@segmentheight,
	database radius/.store in=\database@radius,
	database aspect ratio/.store in=\database@aspectratio,
	database segment height=0.1cm,
	database radius=0.25cm,
	database aspect ratio=0.35,
}
\makeatother

\definecolor{darkgreen}{rgb}{0, 0.625, 0}

%% file: figs/model.tikz
\begin{tikzpicture}
	\node[alice,
	minimum size=.85cm,anchor=south] (User) at (0,0) {};

	\node[database,database radius=0.5cm,database segment height=0.25cm,anchor=south] (DB) at (5,0) {};
	
	\node [below=.2cm of User,align=center] (UserLabel) {User};
	\node [below=.3cm of UserLabel,align=center] (UserOutput) {$\hat{\vec X}^{(M)}(Q,A)$};
	\draw[->] (UserLabel) -- (UserOutput);
	
	\node [below=.2cm of DB,align=center] (DBLabel) {Server};
	\node [below=.3cm of DBLabel,align=center] (DBOutput) {$\hat M(Q)$};
	\draw[->] (DBLabel) -- (DBOutput);
	
	\node [left=.1cm of User.west] {$M$};
	\node [right=.1cm of DB.east] {$\vec X^{[\const{M}]} \begin{cases}
		\vec X^{(1)}\\
		\cdots\\
		\vec X^{(\const M)}
		\end{cases}$};
	
	\path[->] (User.north east) edge[bend left] node[fill=white,anchor=center,pos=0.5] {$Q \sim P_{Q|M}$}  (DB.north west);
	\path[->] (DB.south west) edge[bend left] node[fill=white,anchor=center,pos=0.5] {$A(Q,\vec X^{[\const{M}]})$} (User.south east);
\end{tikzpicture}

%% file: figs/CoverThomas06_1_problem10_5.tex
\begin{tikzpicture}[spy using outlines={circle, magnification=7.5, size=2.5cm, connect spies}]
  
\definecolor{color0}{rgb}{0.75,0,0.75}

\begin{axis}[
width=1\columnwidth,
height=0.3\textheight,
legend cell align={left},
legend style={fill opacity=0.8, draw opacity=1, text opacity=1, draw=white!80!black,
cells={align=left}},
tick pos=both,
x grid style={black!10},
xlabel={Distortion, $\const{D}$},
xmajorgrids,
xmin=0.0, xmax=1.00,
xtick style={color=black},
xtick={0.1,0.2,0.3,0.4,0.5,0.6,0.7,0.8,0.9,1},
y grid style={black!10},
ylabel={Rate},
ylabel near ticks,
xlabel near ticks,
ylabel style={rotate=360},
ymajorgrids,
ymin=0, ymax=15.75,
ytick style={color=black},
ytick={0,2,4,6,8,10,12,14,16},
legend style={font=\footnotesize}
]
\addplot [line width=0.12mm, blue]
table {%
0 12
0.005 11.8493978160974
0.01 11.7188681297382
0.015 11.5959601820962
0.02 11.4780277179764
0.025 11.3638141408317
0.03 11.2525794889269
0.035 11.143834952277
0.04 11.0372332279552
0.045 10.9325147390136
0.05 10.8294780934181
0.055 10.7279624886935
0.06 10.6278365708711
0.065 10.5289910337438
0.07 10.4313335089096
0.075 10.334784923222
0.08 10.2392768318355
0.085 10.1447494205279
0.09 10.0511499796418
0.095 9.95843171816045
0.1 9.86655282812145
0.105 9.77547573660213
0.11 9.68516650050096
0.115 9.59559431158559
0.12 9.50673108778529
0.125 9.41855113272583
0.13 9.33103084983404
0.135 9.24414850050124
0.14 9.15788399813431
0.145 9.07221873167746
0.15 8.98713541351722
0.155 8.9026179477024
0.16 8.81865131519889
0.165 8.73522147351554
0.17 8.65231526852159
0.175 8.56992035666191
0.18 8.48802513608347
0.185 8.40661868543495
0.19 8.32569070930224
0.195 8.24523148940639
0.2 8.16523184082531
0.205 8.08568307261112
0.21 8.00657695226749
0.215 7.92790567362734
0.22 7.84966182773611
0.225 7.77183837639901
0.23 7.69442862809696
0.235 7.61742621601384
0.24 7.540825077951
0.245 7.46461943793291
0.25 7.38880378933178
0.255 7.31337287935968
0.26 7.23832169479481
0.265 7.16364544882355
0.27 7.08933956889369
0.275 7.01539968548581
0.28 6.94182162171979
0.285 6.8686013837226
0.29 6.79573515169124
0.295 6.72321927159139
0.3 6.65105024743866
0.305 6.57922473411464
0.31 6.50773953067434
0.315 6.43659157410647
0.32 6.36577793351107
0.325 6.29529580466272
0.33 6.2251425049306
0.335 6.1553154685289
0.34 6.085812242074
0.345 6.0166304804266
0.35 5.94776794279908
0.355 5.87922248910993
0.36 5.81099207656908
0.365 5.74307475647867
0.37 5.67546867123593
0.375 5.60817205152513
0.38 5.54118321368746
0.385 5.47450055725795
0.39 5.40812256265989
0.395 5.34204778904779
0.4 5.27627487229073
0.405 5.21080252308862
0.41 5.14562952521442
0.415 5.08075473387615
0.42 5.01617707419276
0.425 4.9518955397787
0.43 4.88790919143223
0.435 4.82421715592318
0.44 4.76081862487596
0.445 4.69771285374434
0.45 4.63489916087446
0.455 4.57237692665329
0.46 4.51014559273962
0.465 4.44820466137532
0.47 4.3865536947746
0.475 4.32519231458946
0.48 4.26412020144967
0.485 4.20333709457586
0.49 4.14284279146446
0.495 4.08263714764376
0.5 4.02272007650008
0.505 3.96309154917376
0.51 3.90375159452446
0.515 3.84470029916586
0.52 3.78593780756968
0.525 3.72746432223947
0.53 3.66928010395461
0.535 3.61138547208534
0.54 3.55378080497964
0.545 3.49646654042331
0.55 3.43944317617447
0.555 3.38271127057435
0.56 3.32627144323598
0.565 3.2701243758132
0.57 3.21427081285225
0.575 3.15871156272872
0.58 3.10344749867279
0.585 3.04847955988618
0.59 2.99380875275445
0.595 2.93943615215865
0.6 2.88536290289076
0.605 2.83159022117782
0.61 2.77811939631993
0.615 2.72495179244799
0.62 2.6720888504075
0.625 2.61953208977517
0.63 2.56728311101597
0.635 2.51534359778872
0.64 2.46371531940912
0.645 2.41240013347998
0.65 2.36139998869913
0.655 2.31071692785666
0.66 2.26035309103405
0.665 2.21031071901895
0.67 2.16059215695066
0.675 2.11119985821278
0.68 2.06213638859112
0.685 2.01340443071653
0.69 1.9650067888144
0.695 1.9169463937847
0.7 1.86922630863873
0.705 1.82184973432145
0.71 1.77482001595131
0.715 1.72814064951267
0.72 1.68181528903986
0.725 1.63584775433589
0.73 1.59024203927377
0.735 1.54500232073362
0.74 1.50013296823489
0.745 1.45563855432976
0.75 1.41152386583186
0.755 1.367793915963
0.76 1.32445395751108
0.765 1.28150949710392
0.77 1.23896631071705
0.775 1.1968304605491
0.78 1.1551083134162
0.785 1.11380656083744
0.79 1.07293224100759
0.795 1.03249276288122
0.8 0.992495932625408
0.805 0.952949982736493
0.81 0.913863604162339
0.815 0.875245981825055
0.82 0.837106834003576
0.825 0.799456456112022
0.83 0.762305769501699
0.835 0.725666376025647
0.84 0.689550619239007
0.845 0.653971653272514
0.85 0.618943520617341
0.855 0.584481240307573
0.86 0.550600908294431
0.865 0.517319812191365
0.87 0.484656563054168
0.875 0.452631247475953
0.88 0.421265604065418
0.885 0.390583229395716
0.89 0.360609819841093
0.895 0.331373457472258
0.9 0.302904950521587
0.905 0.275238242090584
0.91 0.248410905101947
0.915 0.222464747518057
0.92 0.197446560355607
0.925 0.173409053272159
0.93 0.150412040489709
0.935 0.128523966853928
0.94 0.107823905511204
0.945 0.0884042248636376
0.95 0.0703742311182456
0.955 0.0538652782437961
0.96 0.0390381687153294
0.965 0.0260942945671392
0.97 0.015293232747009
0.975 0.00698228618182206
0.98 0.0016502648565222
};
\addlegendentry{$R_\textnormal{WPIR+LC}(\const{D},\nicefrac{1}{2})=2\RDF_X(\const{D})$}
\addplot [line width=0.12mm, cyan]
table {%
0 15
0.005 14.7803403907644
0.01 14.5908260203754
0.015 14.4127811827231
0.02 14.2422369349203
0.025 14.0773088300203
0.03 13.9168873814494
0.035 13.7602398386215
0.04 13.6068457201088
0.045 13.4563161422298
0.05 13.3083495076766
0.055 13.1627051127102
0.06 13.0191864358147
0.065 12.8776300368229
0.07 12.7378978922318
0.075 12.5998719316364
0.08 12.4634500375643
0.085 12.3285430492352
0.09 12.1950724737584
0.095 12.062968707539
0.1 11.9321696332058
0.105 11.802619497901
0.11 11.6742680057793
0.115 11.5470695759173
0.12 11.4209827295985
0.125 11.2959695799787
0.13 11.1719954036138
0.135 11.0490282780936
0.14 10.9270387735169
0.145 10.8059996881956
0.15 10.6858858209283
0.155 10.5666737737912
0.16 10.4483417804745
0.165 10.330869556203
0.17 10.2142381659588
0.175 10.0984299083218
0.18 9.98342821269477
0.185 9.86921754805956
0.19 9.75578334170882
0.195 9.64311190664403
0.2 9.53119037653232
0.205 9.42000664728199
0.21 9.30954932443444
0.215 9.19980767568344
0.22 9.09077158793028
0.225 8.98243152836837
0.23 8.87477850914862
0.235 8.76780405524758
0.24 8.66150017519989
0.245 8.55585933440538
0.25 8.4508744307523
0.255 8.34653877233318
0.26 8.24284605705356
0.265 8.1397903539617
0.27 8.03736608614003
0.275 7.93556801502621
0.28 7.83439122603815
0.285 7.73383111539828
0.29 7.63388337805885
0.295 7.53454399664315
0.3 7.43580923132695
0.305 7.33767561059194
0.31 7.24013992279027
0.315 7.1431992084664
0.32 7.04685075338988
0.325 6.95109208225003
0.33 6.8559209529853
0.335 6.76133535170372
0.34 6.66733348816966
0.345 6.5739137918316
0.35 6.48107490836627
0.355 6.38881569671912
0.36 6.29713522663077
0.365 6.20603277662801
0.37 6.1155078324731
0.375 6.02556008606236
0.38 5.93618943476792
0.385 5.84739598121794
0.39 5.75918003351583
0.395 5.67154210589903
0.4 5.58448291984138
0.405 5.49800340560201
0.41 5.41210470423489
0.415 5.32678817006372
0.42 5.24205537364044
0.425 5.15790810520014
0.43 5.074348378635
0.435 4.99137843600682
0.44 4.90900075262496
0.445 4.8272180427219
0.45 4.74603326575208
0.455 4.66544963336316
0.46 4.58547061707395
0.465 4.50609995670841
0.47 4.42734166964322
0.475 4.34920006092615
0.48 4.27167973433514
0.485 4.19478560445172
0.49 4.11852290983871
0.495 4.0428972274133
0.5 3.96791448812745
0.505 3.89358099407544
0.51 3.81990343716489
0.515 3.74688891950755
0.52 3.67454497570367
0.525 3.60287959721854
0.53 3.53190125907534
0.535 3.46161894912354
0.54 3.39204220016972
0.545 3.32318112531227
0.55 3.25504645685745
0.555 3.18764958926422
0.56 3.121002626624
0.565 3.05511843526965
0.57 2.99001070219987
0.575 2.92569400012202
0.58 2.86218386005399
0.585 2.79949685259318
0.59 2.73765067915985
0.595 2.67666427477146
0.6 2.61104631167583
0.605 2.55735339376555
0.61 2.4990740813836
0.615 2.44174518825369
0.62 2.385393916069
0.625 2.33004969472881
0.63 2.26415422017887
0.635 2.22251289525752
0.64 2.15530505383811
0.645 2.11942602183118
0.65 2.04988008563013
0.655 1.99844458495356
0.66 1.97392380457342
0.665 1.8980923411424
0.67 1.84915526232807
0.675 1.80102690578312
0.68 1.75369221831598
0.685 1.70713464346614
0.69 1.66133655809668
0.695 1.61627959035895
0.7 1.57194494512518
0.705 1.52831370969573
0.71 1.48536708018043
0.715 1.4430866588991
0.72 1.40145456094142
0.725 1.36045359058501
0.73 1.32006734009148
0.735 1.28028026164765
0.74 1.24107771752153
0.745 1.20244600403735
0.75 1.16437236239256
0.755 1.12684497736467
0.76 1.08985296499444
0.765 1.05338635623031
0.77 1.01743607683608
0.775 0.981993923619245
0.78 0.947052542476199
0.785 0.912605407276237
0.79 0.878646797412321
0.795 0.84517178180697
0.8 0.81217620337737
0.805 0.77965666727824
0.81 0.747610533936515
0.815 0.716035915259413
0.82 0.684931676099891
0.825 0.654297439341787
0.83 0.624133600220768
0.835 0.59444134166324
0.84 0.565222659574507
0.845 0.536480394802139
0.85 0.508218273281756
0.855 0.4804409561732
0.86 0.453154100155897
0.865 0.426364430809769
0.87 0.400079832282829
0.875 0.374309454511466
0.88 0.349063837991412
0.885 0.324355071340384
0.89 0.300196976542475
0.895 0.276605332845909
0.9 0.25359815303758
0.905 0.231196020319845
0.91 0.209422506548713
0.915 0.18830469335335
0.92 0.167873838445654
0.925 0.148166191983939
0.93 0.129224156500177
0.935 0.111097639004236
0.94 0.0938461057198055
0.945 0.0775411760553009
0.95 0.0622704494276716
0.955 0.0481429852257031
0.96 0.035297610125645
0.965 0.0239161393985872
0.97 0.0142442510973644
0.975 0.00663179173185591
0.98 0.00160619109788751
};
\addlegendentry{$R_{P_2}(\const{D})$}
\addplot [line width=0.12mm, red]
table {%
0 15
0 12
0 12
0.004 11.8769370399674
0.008 11.7699204317074
0.012 11.6689894621405
0.012 11.6689894621405
0.016 11.5720270196315
0.024 11.3863994773052
0.028 11.2967511376047
0.036 11.122352516539
0.04 11.0372332279552
0.04 11.0372332279552
0.044 10.9533181167071
0.048 10.8705019794274
0.052 10.7886972492296
0.052 10.7886972492296
0.056 10.7078299077094
0.06 10.6278365708711
0.064 10.548662355528
0.064 10.548662355528
0.068 10.4702592811255
0.076 10.3156020928841
0.08 10.2392768318355
0.08 10.2392768318355
0.084 10.1635790703483
0.092 10.0139593775866
0.096 9.93999004954896
0.096 9.93999004954896
0.1 9.86655282812145
0.108 9.72120008912498
0.112 9.64925080899054
0.112 9.64925080899054
0.116 9.57776580429749
0.124 9.43613362304505
0.128 9.36596123267081
0.128 9.36596123267081
0.132 9.2962025179061
0.14 9.15788399813431
0.144 9.089304693842
0.148 9.02109996241085
0.152 8.95326138895789
0.152 8.95326138895789
0.156 8.8857810173666
0.16 8.81865131519889
0.164 8.75186514212182
0.176 8.55350169616213
0.18 8.48802513608347
0.184 8.42286137675432
0.188 8.35800511717937
0.188 8.35800511717937
0.192 8.29345129645568
0.2 8.16523184082531
0.204 8.10155715731772
0.204 8.10155715731772
0.208 8.03816679145266
0.212 7.97505668330991
0.216 7.91222294023113
0.216 7.91222294023113
0.22 7.84966182773611
0.228 7.7253432975609
0.232 7.66357912901875
0.232 7.66357912901875
0.236 7.60207407532128
0.244 7.47982919416082
0.248 7.41908359149492
0.248 7.41908359149492
0.252 7.35858554266773
0.26 7.23832169479481
0.264 7.17855092540506
0.264 7.17855092540506
0.268 7.11901776101307
0.276 7.0006552575419
0.28 6.94182162171979
0.28 6.94182162171979
0.284 6.88321699104895
0.292 6.76668695665533
0.296 6.70875782745753
0.296 6.70875782745753
0.3 6.65105024743866
0.308 6.53629297391716
0.312 6.47924004545201
0.312 6.47924004545201
0.316 6.42240219223572
0.324 6.30936583997063
0.328 6.253164532104
0.328 6.253164532104
0.332 6.19717267863157
0.34 6.085812242074
0.344 6.03044122460185
0.344 6.03044122460185
0.348 5.97527479014792
0.352 5.92031182779101
0.356 5.86555126700359
0.356 5.86555126700359
0.36 5.81099207656908
0.368 5.70247387233371
0.372 5.64851298366415
0.372 5.64851298366415
0.552 3.22799851232375
0.556 3.17425980394552
0.56 3.12100262662395
0.572 2.9641882502093
0.576 2.91292693240515
0.58 2.86218386005401
0.584 2.81196757088401
0.584 2.81196757088401
0.588 2.76228706552709
0.592 2.71315184074476
0.596 2.66457192605722
0.608 2.52227320985388
0.612 2.47602699763169
0.612 2.47602699763169
0.616 2.43039581602801
0.656 1.98825868915065
0.664 1.90797803625017
0.664 1.90797803625017
0.668 1.86863223812923
0.676 1.79149698891723
0.68 1.75369225152529
0.7 1.57194494812127
0.704 1.53698458507107
0.712 1.46837598238344
0.716 1.43470893437138
0.716 1.43470893437138
0.72 1.40145456094286
0.724 1.36860409405651
0.728 1.3361490889568
0.736 1.27239344217961
0.74 1.24107771756009
0.74 1.24107771756009
0.744 1.21012730632343
0.748 1.17953562760457
0.752 1.14929648790768
0.752 1.14929648790768
0.756 1.11940407703161
0.76 1.08985296498127
0.764 1.06063809404828
0.764 1.06063809404828
0.768 1.03175477330917
0.776 0.974965806498162
0.78 0.947052542572426
0.78 0.947052542572426
0.784 0.919455580757211
0.788 0.892171951340568
0.792 0.86519900981898
0.792 0.86519900981898
0.796 0.838534430613109
0.804 0.786122628686487
0.808 0.760372316638666
0.808 0.760372316638666
0.812 0.734924185348973
0.82 0.684931676051323
0.824 0.660386678745456
0.824 0.660386678745456
0.828 0.636142630447207
0.836 0.588559640678995
0.84 0.565222659421813
0.84 0.565222659421813
0.844 0.542190568966562
0.852 0.49704887621191
0.856 0.474944148681287
0.856 0.474944148681287
0.86 0.453154100197839
0.864 0.43168222843257
0.868 0.410532507315784
0.868 0.410532507315784
0.872 0.389709418060315
0.88 0.349063837694595
0.884 0.32925324009231
0.884 0.32925324009231
0.888 0.309793170777075
0.896 0.271956520608001
0.9 0.253598153047279
0.908 0.218054830191966
0.912 0.200895048582948
0.912 0.200895048582948
0.916 0.18416248302479
0.924 0.152047930901235
0.928 0.136706087321668
0.936 0.107575236170032
0.94 0.0938461046051606
0.944 0.0807224959881068
0.948 0.0682482157590911
0.948 0.0682482157590911
0.952 0.0564752757002371
0.956 0.0454662516878062
0.96 0.035297610125645
0.96 0.035297610125645
0.964 0.0260645542306173
0.968 0.0178883707424273
0.972 0.0109277649900279
0.984 1.30877356152581e-05
0.984 1.30538830172e-05
};
\addlegendentry{$\conv{2\RDF_X}{R_{P_2}}(\const{D})$}
\end{axis}
\spy [red] on (3.5,1.55) in node [left] at (7.25,3); 
\end{tikzpicture}

%% file: figs/M16beta20L1o16.tikz
\begin{tikzpicture}[font=\scriptsize]

\definecolor{mplblue}{rgb}{0.12156862745098,0.466666666666667,0.705882352941177}
\definecolor{mplorange}{rgb}{1,0.498039215686275,0.0549019607843137}
\definecolor{mplgreen}{rgb}{0.172549019607843,0.627450980392157,0.172549019607843}
\definecolor{mplred}{rgb}{0.83921568627451,0.152941176470588,0.156862745098039}
\definecolor{mplpurple}{rgb}{0.580392156862745,0.403921568627451,0.741176470588235}
\definecolor{mplbrown}{rgb}{0.549019607843137,0.337254901960784,0.294117647058824}

\begin{axis}[
width=0.36\textwidth,
height=0.22\textheight,
legend cell align={left},
legend style={fill opacity=0.8, draw opacity=1, text opacity=1, draw=white!80!black,
cells={align=left}},
tick pos=both,
x grid style={black!10},
xlabel={Distortion, $\const D$},
xmajorgrids,
xmin=0, xmax=0.51,
xtick style={color=black},
xtick={0,0.125,0.25,0.375,0.5},
y grid style={black!10},
ylabel={Rate},
ylabel style={rotate=360, left=-5mm},
ymajorgrids,
ymin=0, ymax=16.8,
ytick style={color=black},
ytick={0,4,8,12,16},
legend style={font=\tiny}
]


\addplot[dashed,semithick,mplblue] 
table[x=D,y=R] {
	D      R 
	0      16 
	0.125  10.21136 
	0.1875 7.829904 
	0.3125 3.59104 
	0.5    0 
};
\addlegendentry{$R_{\mathrm{rep}}(\const{D},\nicefrac{1}{16})$};

\addplot[semithick,mplblue] 
table[x=D,y=R] {
	D      R 
   0.000000  16.0
   0.137296   8.8
   0.154355   8.0
   0.173079   7.2
   0.193832   6.4
   0.216521   5.6
   0.241560   4.8
   0.268502   4.0
   0.298352   3.2
   0.366300   1.6
   0.411901   0.8
   0.500000   0.0
};
\addlegendentry{$R_{\mathrm{rnd}}(\const{D},\nicefrac{1}{16})$};

\addplot [thick,dash pattern=on 1pt off 3pt on 3pt off 3pt,mplblue] 
table[x=D,y=R] {
    D     R
0.00676676	16.8
0.018394	16.
0.0303279	15.2
0.0392024	14.4
0.0477467	13.6
0.0604279	12.8
0.0744695	12.
0.0870877	11.2
0.101662	10.4
0.118049	9.6
0.134571	8.8
0.153095	8.
0.172817	7.2
0.194369	6.4
0.217869	5.6
0.243788	4.8
0.272621	4.
0.305117	3.2
0.342316	2.4
0.385691	1.6
0.437315	0.8
0.5	0.
};
\addlegendentry{$R_{\mathrm{KV}}(\const{D},\nicefrac{1}{16})$};

\addplot [thick,dotted,mplblue] 
table[x=D,y=R] {
	D     R 
0.	16.
0.025	13.3014
0.05	11.4176
0.075	9.85102
0.1	8.49607
0.125	7.30297
0.15	6.24256
0.175	5.29575
0.2	4.44915
0.225	3.69291
0.25	3.01955
0.275	2.42323
0.3	1.89935
0.325	1.44422
0.35	1.05491
0.375	0.729056
0.4	0.46479
0.425	0.260668
0.45	0.115609
0.475	0.0288659
0.5	0.
};
\addlegendentry{$R^*(\const{D},\nicefrac{1}{16})$};

\end{axis}

\end{tikzpicture}

%% file: figs/M16beta20L1o8.tikz
\begin{tikzpicture}[font=\scriptsize]

\definecolor{mplblue}{rgb}{0.12156862745098,0.466666666666667,0.705882352941177}
\definecolor{mplorange}{rgb}{1,0.498039215686275,0.0549019607843137}
\definecolor{mplgreen}{rgb}{0.172549019607843,0.627450980392157,0.172549019607843}
\definecolor{mplred}{rgb}{0.83921568627451,0.152941176470588,0.156862745098039}
\definecolor{mplpurple}{rgb}{0.580392156862745,0.403921568627451,0.741176470588235}
\definecolor{mplbrown}{rgb}{0.549019607843137,0.337254901960784,0.294117647058824}

\begin{axis}[
width=0.36\textwidth,
height=0.22\textheight,
legend cell align={left},
legend style={fill opacity=0.8, draw opacity=1, text opacity=1, draw=white!80!black,
cells={align=left}},
tick pos=both,
x grid style={black!10},
xlabel={Distortion, $\const D$},
xmajorgrids,
xmin=0, xmax=0.51,
xtick style={color=black},
xtick={0,0.125,0.25,0.375,0.5},
y grid style={black!10},
ymajorgrids,
ymin=0, ymax=8.4,
ytick style={color=black},
ytick={0,2,4,6,8},
legend style={font=\tiny}
]


\addplot[dashed,semithick,mplorange] 
table[x=D,y=R] {
	D      R 
	0      8 
	0.125  5.10568 
	0.1875 3.91488 
	0.3125 1.79552 
	0.5    0 
};
\addlegendentry{$R_{\mathrm{rep}}(\const{D},\nicefrac{1}{8})$};

\addplot[semithick,mplorange] 
table[x=D,y=R] {
	D      R 
   0.000000  8.0
   0.137296  4.4
   0.154355  4.0
   0.173079  3.6
   0.193832  3.2
   0.216521  2.8
   0.241560  2.4
   0.268502  2.0
   0.298352  1.6
   0.366300  0.8
   0.411901  0.4
   0.500000  0.0
};
\addlegendentry{$R_{\mathrm{rnd}}(\const{D},\nicefrac{1}{8})$};

\addplot [thick,dash pattern=on 1pt off 3pt on 3pt off 3pt,mplorange] 
table[x=D,y=R] {
    D     R
0.00676676	8.4
0.018394	8.
0.0303279	7.6
0.0392024	7.2
0.0477467	6.8
0.0604279	6.4
0.0744695	6.
0.0870877	5.6
0.101662	5.2
0.118049	4.8
0.134571	4.4
0.153095	4.
0.172817	3.6
0.194369	3.2
0.217869	2.8
0.243788	2.4
0.272621	2.
0.305117	1.6
0.342316	1.2
0.385691	0.8
0.437315	0.4
0.5	0.
};
\addlegendentry{$R_{\mathrm{KV}}(\const{D},\nicefrac{1}{8})$};

\addplot [thick,dotted,mplorange] 
table[x=D,y=R] {
	D     R 
0.	8.
0.025	6.65071
0.05	5.70882
0.075	4.92551
0.1	4.24804
0.125	3.65148
0.15	3.12128
0.175	2.64787
0.2	2.22458
0.225	1.84646
0.25	1.50978
0.275	1.21161
0.3	0.949673
0.325	0.722111
0.35	0.527456
0.375	0.364528
0.4	0.232395
0.425	0.130334
0.45	0.0578044
0.475	0.014433
0.5	0
};
\addlegendentry{$R^*(\const{D},\nicefrac{1}{8})$};

\end{axis}

\end{tikzpicture}

%% file: figs/M16beta20L1.tikz
\begin{tikzpicture}[font=\scriptsize]

\definecolor{mplblue}{rgb}{0.12156862745098,0.466666666666667,0.705882352941177}
\definecolor{mplorange}{rgb}{1,0.498039215686275,0.0549019607843137}
\definecolor{mplgreen}{rgb}{0.172549019607843,0.627450980392157,0.172549019607843}
\definecolor{mplred}{rgb}{0.83921568627451,0.152941176470588,0.156862745098039}
\definecolor{mplpurple}{rgb}{0.580392156862745,0.403921568627451,0.741176470588235}
\definecolor{mplbrown}{rgb}{0.549019607843137,0.337254901960784,0.294117647058824}

\begin{axis}[
width=0.35\textwidth,
height=0.22\textheight,
legend cell align={left},
legend style={fill opacity=0.8, draw opacity=1, text opacity=1, draw=white!80!black,
cells={align=left}},
tick pos=both,
x grid style={black!10},
xlabel={Distortion, $\const D$},
xmajorgrids,
xmin=0, xmax=0.51,
xtick style={color=black},
xtick={0,0.125,0.25,0.375,0.5},
y grid style={black!10},
ymajorgrids,
ymin=0, ymax=1.05,
ytick style={color=black},
ytick={0,0.25,0.5,0.75,1},
legend style={font=\tiny}
]


\addplot[dashed,semithick,mplgreen] 
table[x=D,y=R] {
	D      R 
	0      1 
	0.125  0.63821 
	0.1875 0.48937 
	0.3125 0.22444 
	0.5    0 
};
\addlegendentry{$R_{\mathrm{rep}}(\const{D},1)$};

\addplot[semithick,mplgreen] 
table[x=D,y=R] {
	D			R 
   0.000000  1.00
   0.137296  0.55
   0.154355  0.50
   0.173079  0.45
   0.193832  0.40
   0.216521  0.35
   0.241560  0.30
   0.268502  0.25
   0.298352  0.20
   0.366300  0.10
   0.411901  0.05
   0.500000  0.00
};
\addlegendentry{$R_{\mathrm{rnd}}(\const{D},1)$};

\addplot [thick,dash pattern=on 1pt off 3pt on 3pt off 3pt,mplgreen] 
table[x=D,y=R] {
    D     R
0.00676676	1.05
0.018394	1.
0.0303279	0.95
0.0392024	0.9
0.0477467	0.85
0.0604279	0.8
0.0744695	0.75
0.0870877	0.7
0.101662	0.65
0.118049	0.6
0.134571	0.55
0.153095	0.5
0.172817	0.45
0.194369	0.4
0.217869	0.35
0.243788	0.3
0.272621	0.25
0.305117	0.2
0.342316	0.15
0.385691	0.1
0.437315	0.05
0.5	0.
};
\addlegendentry{$R_{\mathrm{KV}}(\const{D},1)$};

\addplot [thick,dotted,mplgreen] 
table[x=D,y=R] {
	D     R 
0.	1.
0.025	0.831339
0.05	0.713603
0.075	0.615688
0.1	0.531004
0.125	0.456436
0.15	0.39016
0.175	0.330984
0.2	0.278072
0.225	0.230807
0.25	0.188722
0.275	0.151452
0.3	0.118709
0.325	0.0902639
0.35	0.0659319
0.375	0.045566
0.4	0.0290494
0.425	0.0162917
0.45	0.00722555
0.475	0.00180412
0.5	0
};
\addlegendentry{$R^*(\const{D},1)$};

\end{axis}

\end{tikzpicture}